\documentclass{amsart}

\usepackage{amssymb,epsfig,amsmath}
\usepackage{hyperref} 
\usepackage{graphicx}
\usepackage[caption=false]{subfig}
\graphicspath{{./Figures/}}
\usepackage{url}
\usepackage{booktabs}
\usepackage[official]{eurosym}
\usepackage{color}
\usepackage{bm}
\usepackage{enumerate}

\newtheorem{prop}{Proposition}
\newtheorem{cor}{Corollary}
\newtheorem{lem}{Lemma}

\newtheorem{deff}{Definition}
\newtheorem{example}{{\it Example}}
\newtheorem{assumption}{{\it Assumption}}

\newtheorem{remark}{{\it Remark}}

\newcommand{\E}{\mathbb{E}}
\newcommand{\Id}{\mathrm{d}}
\newcommand{\R}{\mathbb{R}}

\title{Risk-Neutral Pricing of Financial Instruments in Emission Markets: A Structural Approach}
\author{Sam Howison}
\address{Mathematical Institute, University of Oxford and
  Oxford-Man Institute, 0X26ED Oxford, UK.}
  \email{howison@maths.ox.ac.uk}
\author{Daniel Schwarz}
\address{Carnegie Mellon University, Department of
  Mathematical Sciences, 5000 Forbes Avenue, Pittsburgh, PA,
  15213-3890, USA.}
  \email{schwarzd@andrew.cmu.edu}
\thanks{The second named author acknowledges support from the Carnegie
  Mellon-Portugal Program, grant UTA\_CMU/MAT/0006/2009 (FCT).}

\begin{document}
\begin{abstract}
We present a novel approach to the pricing of financial instruments in emission markets---for example, the European
Union Emissions Trading Scheme (EU ETS).
The proposed structural model is positioned between existing complex full equilibrium models and pure reduced-form
models. Using an exogenously specified demand for a polluting good, it gives a causal explanation for the accumulation
of CO$_2$ emissions and takes into account the feedback effect from the cost of carbon to the rate at which the
market emits CO$_2$. We derive a forward-backward stochastic differential equation for the price process of the
allowance certificate and solve the associated semilinear partial differential equation numerically. We also show
that derivatives written on the allowance certificate satisfy a linear partial differential equation. The model
is extended to emission markets with multiple compliance periods, and we analyze the impact different intertemporal
connecting mechanisms, such as borrowing, banking, and withdrawal, have on the allowance price.
\end{abstract}

\maketitle

%




\pagestyle{myheadings}
\thispagestyle{plain}
\markboth{SAM HOWISON AND DANIEL SCHWARZ}{RISK-NEUTRAL PRICING IN EMISSION MARKETS}

\section{Introduction}
Global warming has been recognized by policy makers as a key 21st century problem. The phenomenon is widely believed to be the result of a greenhouse
effect that is caused by increases in atmospheric gases such as carbon dioxide, methane, ozone, and water vapor. Forced to address this issue, 37
countries ratified the Kyoto Protocol on December 11, 1997 in Kyoto, Japan. Under this agreement, binding limits, expressed in assigned amount units
(AAUs) and measured in tonnes of CO$_2$ equivalent greenhouse gas (GHG),\footnote{The CO$_2$ equivalent of a given greenhouse gas denotes the amount of
CO$_2$ that has the same global warming potential over a specified timescale.} are imposed on the emissions of participating countries. To meet their
obligations, countries may draw upon---amongst other mechanisms---any of the following three market-based mechanisms:
\begin{enumerate}
 \item The \textit{Clean Development Mechanism} (CDM), defined in article 12 of the Kyoto Protocol, allows countries to implement emission-reduction
projects in developing countries. For this they receive certified emission reduction (CER) credits, each worth one tonne of CO$_2$ equivalent, which
can be used for meeting Kyoto targets.
 \item The \textit{Joint Implementation} (JI) mechanism, defined in article 6 of the Kyoto Protocol, allows countries to earn emission reduction units
(ERUs), each worth one tonne of CO$_2$ equivalent, from establishing emission-reduction projects in other Kyoto countries. Like CERs, these units can
be used to meet Kyoto targets.
 \item \textit{Emissions Trading}, as defined in article 17 of the Kyoto Protocol, allows market participants that have AAUs, CERs, or ERUs to spare, to
sell their excess capacity to other participants. This creates the so-called carbon market.
\end{enumerate}

The Kyoto Protocol merely constitutes a global framework that encourages participating countries to put in place platforms on which CERs, ERUs, and
AAUs can be traded. Subject to broad provisions, the market design of any local implementation of an emissions trading system is left to the hosting
countries.

In this paper we present a simple model for emissions trading. Our model incorporates market design features which are commonly found in successful
implementations of such trading schemes---for example, the SO$_2$ and NOx trading programs in the US or the European Union Emissions Trading Scheme (EU
ETS). Because of its pioneering role, we choose the latter example to illustrate the working principle of emissions trading.

\subsection{Emissions trading in the EU ETS}
The limit on emissions during one \textit{compliance period}, also referred
to as the cap on emissions, is realized through an \textit{initial
  allocation} of \textit{allowance certificates}---each worth one EU
allowance unit (EUA) and permitting its holder to emit one tonne of CO$_2$
equivalent\footnote{For simplicity we write CO$_2$ from now on whenever we
  mean CO$_2$ equivalent GHGs.} GHGs---by the government to firms in the
market. At the end of each compliance period, firms must offset their
accumulated emissions by submitting an adequate number of certificates. If
they fail to do so (the event of \textit{noncompliance}), they must pay a
monetary \textit{penalty} for each unit of excess emissions. Throughout a
compliance period allowances are traded actively, and this leads to the
formation of a price, which represents the cost of carbon. Firms can then buy allowances to avoid the penalty,
or exploit their own pollution-light production by selling them. 

In practice, an emissions trading scheme consists of multiple compliance periods, each with its own distinct cap and penalty. Subsequent periods are
joined by connecting mechanisms, which regulate the transition from one compliance period to the next. The key mechanisms go by the names of
\textit{banking}, \textit{borrowing}, and \textit{withdrawal}. The banking mechanism allows market participants to carry allowance certificates,
allocated for compliance at the end of the current period, forward to the next compliance period. Similarly, borrowing enables firms to use the next
period's certificates for compliance at the end of the current trading period. The withdrawal mechanism constitutes additional punishment for
noncompliance: it prescribes that, in addition to the monetary penalty payment, one allowance certificate from the next period's allocation is withdrawn
for each unit of excess emissions at the end of the current period.

Since the Linking Directive came into force, the EU has been accepting credits from the CDM and the JI mechanism
for compliance in its trading scheme. Because one EUA is equivalent to an AAU, the base unit of the Kyoto Protocol, CERs, ERUs, and EUAs can all be
traded within the same system straightforwardly. In practice, this takes place mostly on platforms such as the European Energy Exchange (EEX), where
EUA and CER spot and future contracts are traded actively.

Emission reduction as part of a trading scheme occurs in two ways. The immediate consequence is to shift production within the existing fleet of
resources to pollution friendlier ones---an effect we refer to as \textit{load shifting}. The cost of carbon also makes it attractive for firms to
invest in \textit{long-term abatement measures} if the cost of reducing their emissions by one unit lies below the value of an allowance certificate.
Even if a firm has sufficient allowances to cover its emissions, it should make use of all available emission-reduction measures whose marginal
abatement cost (MAC) lies below the value of the allowance certificate. It can then sell spare certificates to companies whose MAC is above the market
price of allowances and make a profit. For this reason it has been argued that cap-and-trade schemes provide emission reduction at the lowest cost to
society. However, there is also evidence which suggests that the implied cost of carbon to make long-term investment in renewables such as solar cells
worthwhile is \$196 per tonne of CO$_2$; this is far above current allowance prices (cf. \cite{economist2009}).

\subsection{Electricity generation: A pollution-intensive process}\label{str:intro_el}
The primary process that releases CO$_2$ emissions is the burning of fossil fuels. Since this is heavily used for the generation of electric power,
electricity offers itself as an exemplary good for the academic study of emission markets. A wide spectrum of technologies, including nuclear fission,
wind turbines, hydropower, and the burning of fossil fuels, are used for the generation of electricity. Because these technologies differ substantially
in their emission rates it is important to identify which generators are used in the market at any point in time. In principle, this can be deduced
from the electricity bid stack.

The \textit{bid stack}, introduced in \cite{mBarlow2002,jHinz2003,rAid2009,mCoulon2009a}, aggregates the bidding behavior of firms that supply
electricity. A bid is the amount of electricity a single generator is willing to supply at a specific price. Firms submit their Pareto-optimal bids
for each hour of the next trading day to a central market administrator. An example would be a generator submitting bids (600MW,100\euro),
(200MW,120\euro), and (200MW,200\euro). This generator offers to sell its first 600MW for the specified hour at a price of 100\euro, the next 200MW at a
price of 120\euro, and a further 200MW at a price of 200\euro. Consequently, each firm submits an increasing simple (step) function that maps
electricity supply to its marginal price. The market administrator aggregates the bids for each price level and arranges them in increasing order of
price. Using the cheapest bids first, electricity is supplied at the marginal price of the last unit of electricity that is needed in order to satisfy
demand.

The bids of generators reflect their production costs (cf. \cite{mCoulon2009a}). In particular, firms consider fixed and variable costs when deciding
upon their bid levels. In the absence of emissions trading---a scenario called \textit{business-as-usual}---the latter costs consist predominantly
of the price to be paid by a particular plant for the amount of fuel necessary to generate each unit of electricity (the plant's heat rate multiplied
by the price of the utilized fuel). The introduction of a cap-and-trade system levies a cost on emissions. In principle, firms may remain idle and
sell unused certificates to the market. Therefore, if they choose to produce and to utilize their certificates for compliance purposes, this forgone
profit constitutes an opportunity cost, which leads to an increase in variable costs. As a consequence bids increase by an amount equal to the
marginal emissions rate of the plant (measured in tonnes of CO$_2$ per MWh) multiplied by the allowance price.

Although not a requirement for the model that we propose, in the absence of emissions trading, pollution-intensive fuels have historically gathered on
the left end of the bid stack because they are cheaper to use, whereas environmentally friendlier technologies tend to be more expensive and are
concentrated further to the right.\footnote{An exception to this rule of thumb constitutes must-run bids, which are always placed on the left end of
the bid stack and may, for example, contain bids from nuclear generators that do not emit.} The rationale behind cap-and-trade is that for
sufficiently high carbon costs, pollution-intensive technologies become more expensive than environmentally friendlier ones. The market administrator
\textit{rearranges} bids to preserve the increasing order, and as a result environmentally friendly technologies are now called upon before pollution
intensive ones, leading to cleaner production of electricity.

\begin{figure}[htbp]
  \centering
  \subfloat{\label{fig:ex1_bids}\includegraphics[width=0.3\textwidth]{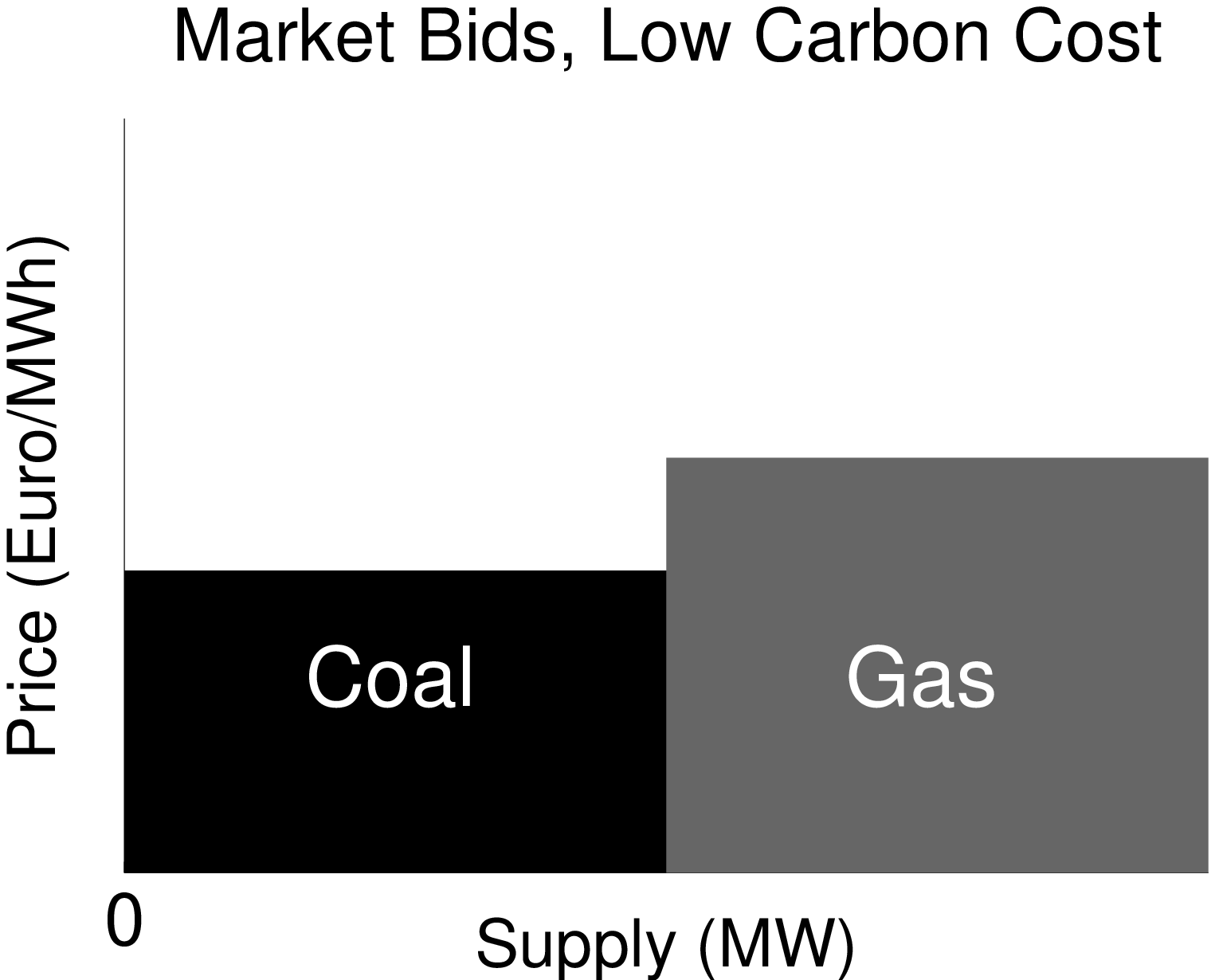}}
  \subfloat{\label{fig:ex1_bids_all}\includegraphics[width=0.3\textwidth]{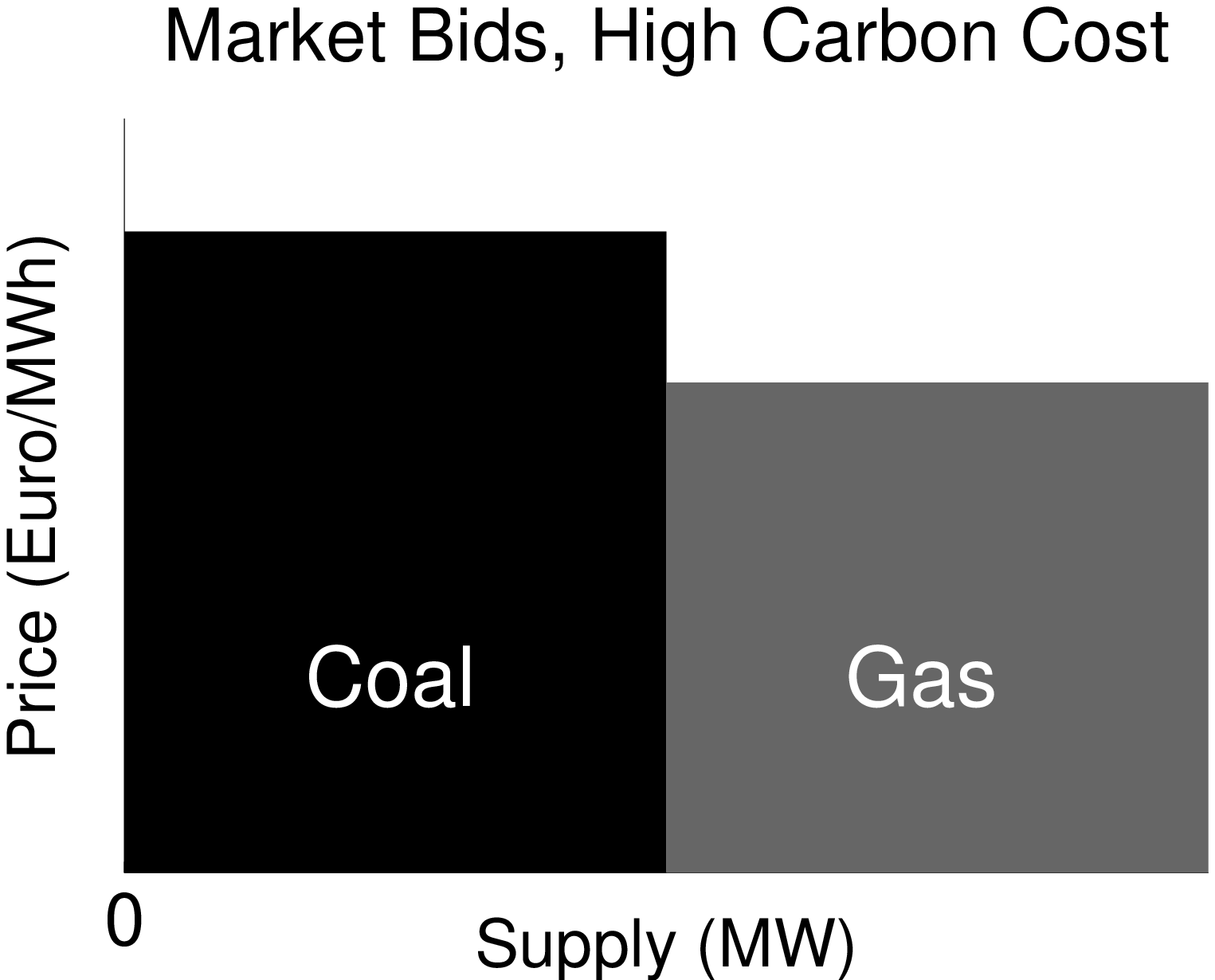}}
  \subfloat{\label{fig:ex1_bids_rearr}\includegraphics[width=0.3\textwidth]{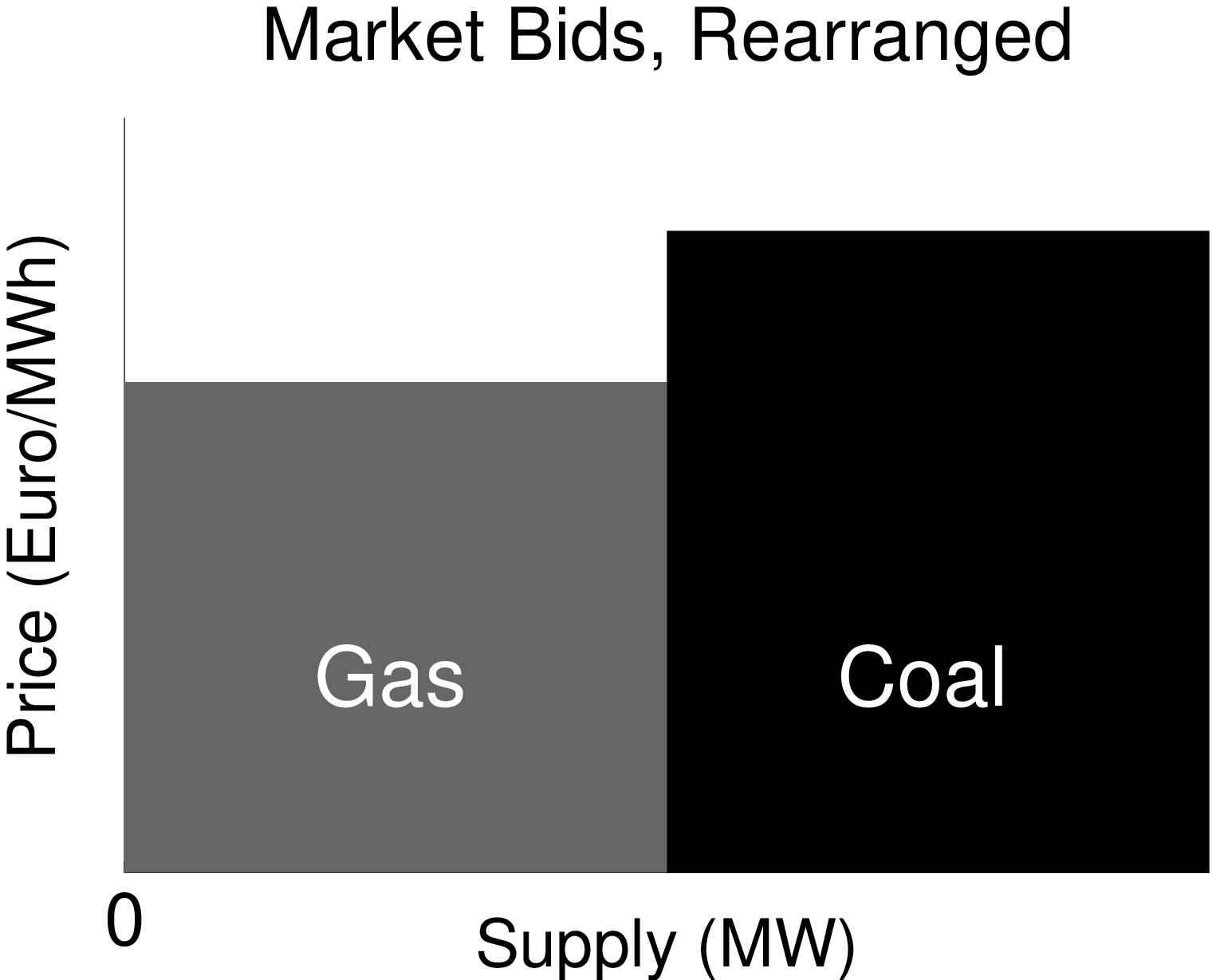}}\\
  \subfloat{\label{fig:ex1_ems}\includegraphics[width=0.3\textwidth]{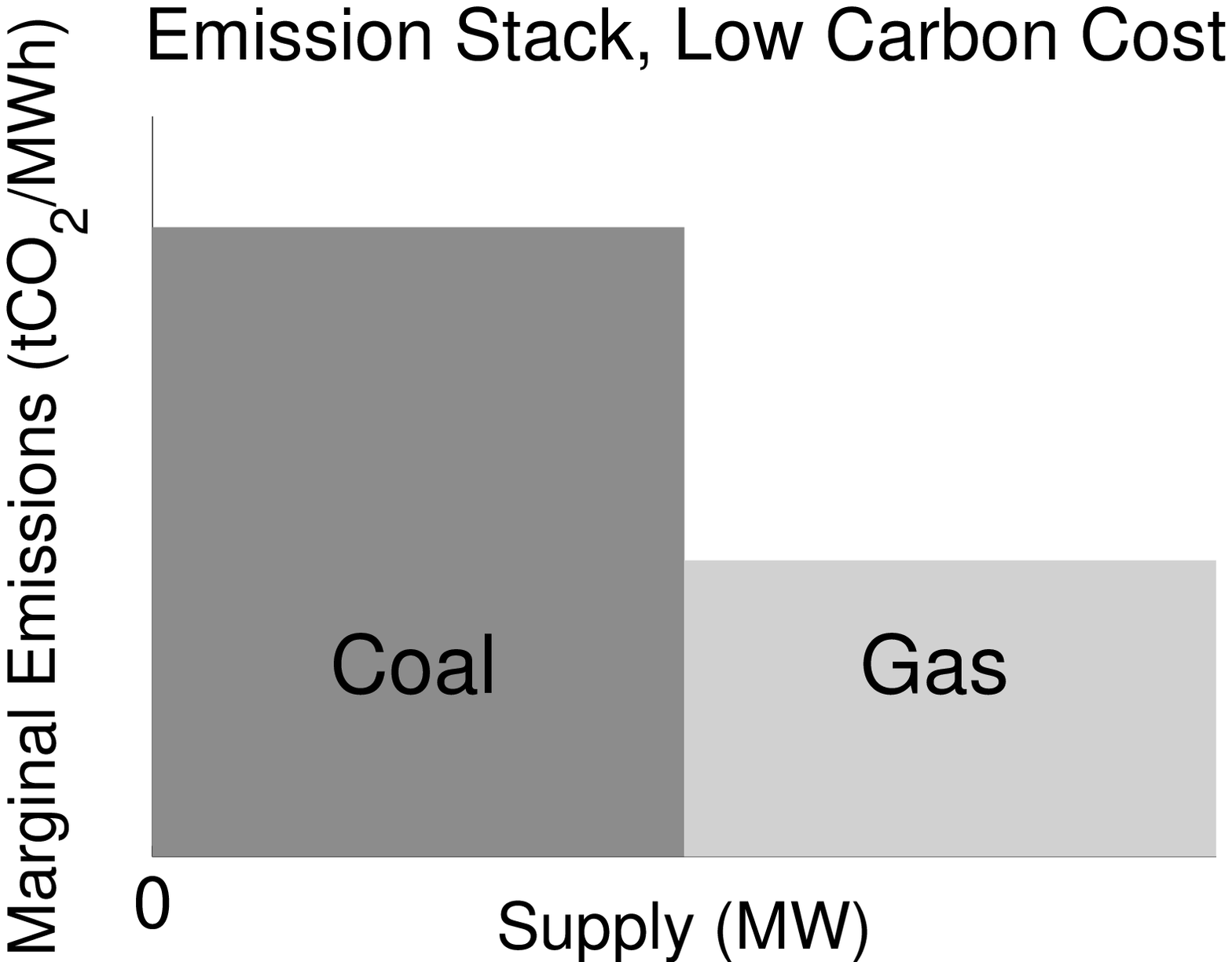}}
  \subfloat{\label{fig:ex1_ems_all}\includegraphics[width=0.3\textwidth]{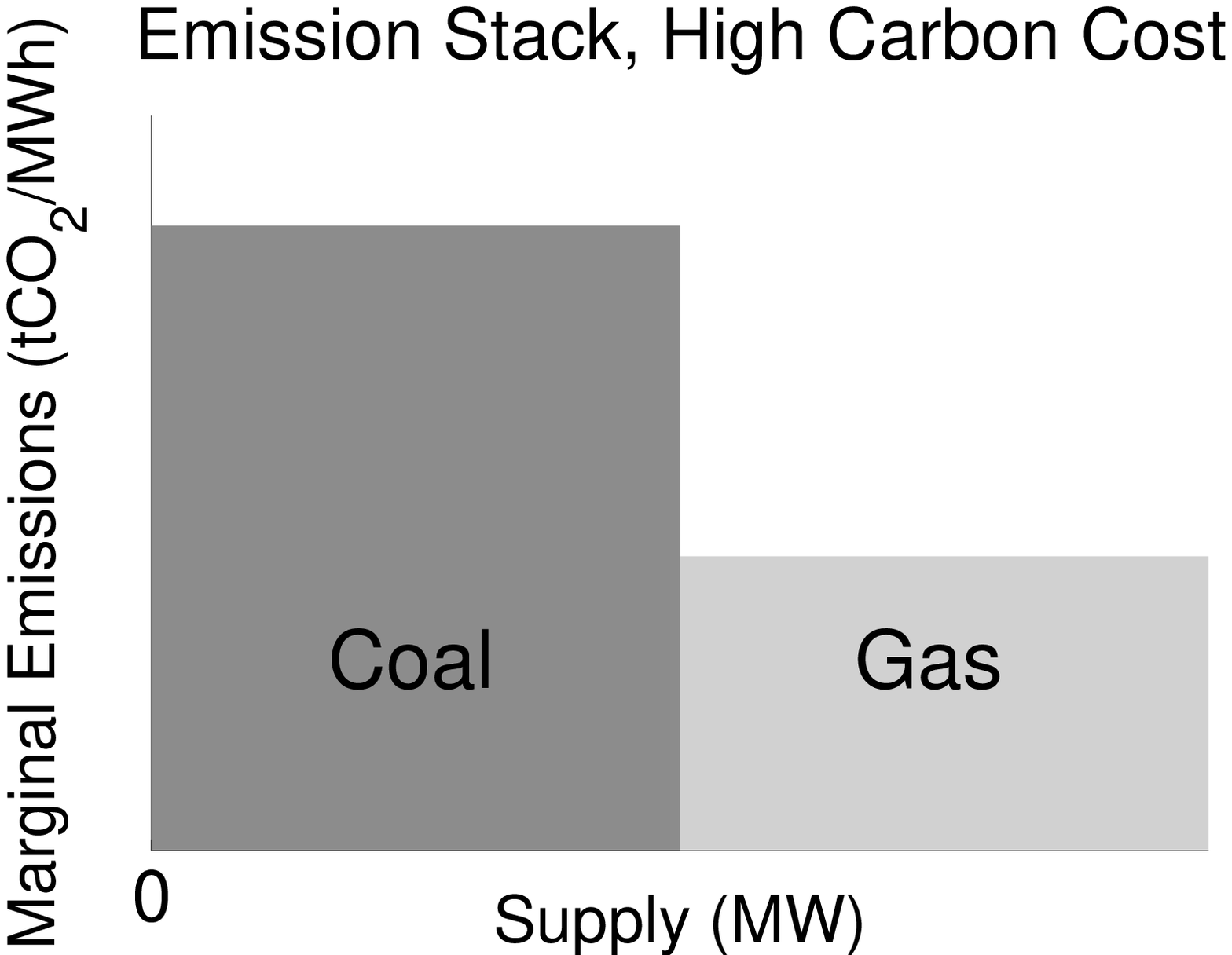}}
  \subfloat{\label{fig:ex1_ems_rearr}\includegraphics[width=0.3\textwidth]{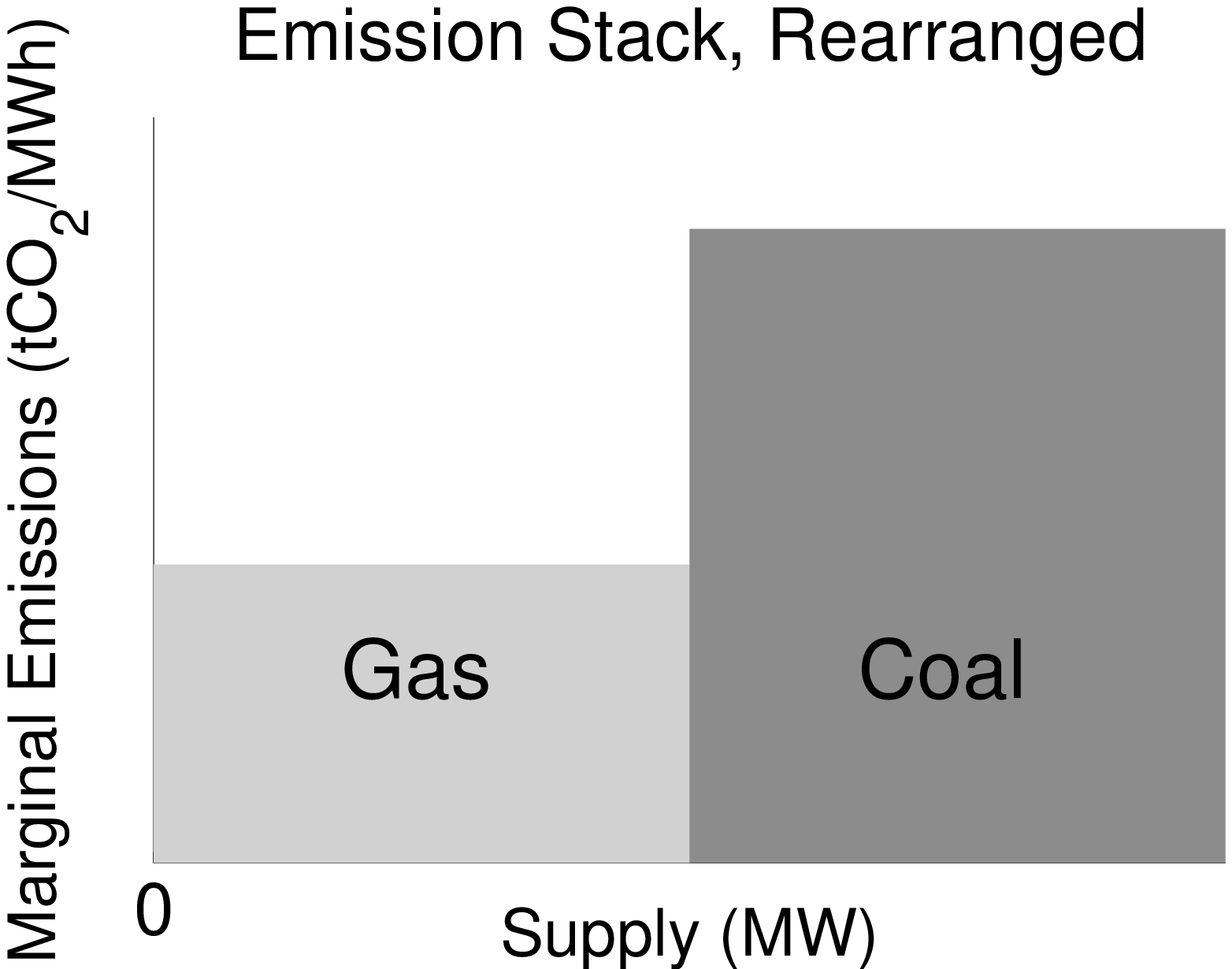}}
  \caption{A schematic of the rearrangement of the bid and the emissions stacks as the cost of carbon increases.}
  \label{fig:ex1_stacks_rearr}
\end{figure}

\begin{example}\rm Consider a simple market with one coal generator and one gas generator, who bid at only one price level.
  We illustrate the influence of emissions trading on the bid stack in Figure
  \ref{fig:ex1_stacks_rearr}.  Initially the cost of carbon is low, and bids
  from coal generators are cheaper than those from gas generators.
  Accordingly, coal bids come first in the bid stack, and the marginal
  emissions corresponding to bids further to the left in the stack are
  relatively higher than those corresponding to bids on the right. As
  emissions become more costly the bid levels from coal and gas generators
  increase, with coal bids doing so more than gas bids. This results in the
  market administrator rearranging the bid stack and placing gas first. The
  result of the higher allowance price is lower emissions, as intended.
\end{example}

The price setting mechanism described above applies directly to day-ahead spot prices set by uniform auctions, as is the case at most exchanges today.
For example, the power spot price for Germany, Austria, Switzerland, and France is determined by such auctions organized by the EEX. Although, since
the onset of electricity market deregulation in 1998, the auction-based trading volume at the EEX has increased substantially over the last few 
years---from 49 TWh in 2003 to 279 TWh in 2010 (cf.~\cite{EEX2011})---a large share of electricity in Europe is still traded over-the-counter or on a
forward basis. However, we believe that in a competitive market with rational agents the day-ahead auction price also serves as the key reference
point for real-time and over-the-counter prices (cf.~\cite{aOckenfels2008}).

\subsection{Literature review}
The first academic treatment of emission markets can be traced back to \cite{jDales1968,dMontgomery1972}. Early models of allowance trading in
discrete and in continuous time were proposed in \cite{mCronshaw1996,pLeiby2001,aMaeda2004,jRubin1996,sSchennach2000,tTietenberg2006}. More recently,
emission markets have been treated from two different angles. On the one hand are full equilibrium models that derive the price processes of
allowances and goods (the production of which causes pollution) from the preferences of individual firms and additional sources of uncertainty. These
have proved insightful but rather cumbersome in their complexity
(cf. \cite{rCarmona2010f,rCarmona2009}). On the other hand are
approaches, which rely on the concept of absence of arbitrage
(i.e. ruling out the possibility of making a profit starting from nothing) to
specify the allowance price evolution directly as the expectation of
the discounted future cash flows under a probability structure which,
in the mathematical finance literature, is called risk-neutral (cf. \cite{jHarrison1981}); then the parameters in the model are calibrated to market data. In these models, the event of
noncompliance is described exogenously, and no causal explanation is given for the accumulation of emissions in the economy. Within this class of
models one can distinguish between models that ignore the feedback from the allowance price to the rate at which firms emit (cf. \cite{rCarmona2010d})
and those that take this feedback effect into account through an
exogenously specified abatement function (cf. \cite{kBorovkov2010,rCarmona2010g,jHinz2010}).

\subsection{The current paper} 
In this paper we propose a structural model which draws upon elements of the
equilibrium approach but still retains the simplicity of the risk-neutral
approach. We take as a starting point an exogenously specified stochastic
process representing demand for electricity,\footnote{Although we have
  formulated the problem in terms of electricity generation (which is,
  indeed, responsible for a large proportion of the emissions covered by the
  Kyoto Protocol), we may at least conceptually extend our model to all
  emissions covered by Kyoto if we view the formation of the equilibrium
  price for energy as equivalent to the action of the market regulator's
  arrangement of the bid stack in increasing price order.} 
and we regard
allowances as derivatives (that is, contingent claims: securities whose
value at a specified future date is determined by the state of the world at
that time, but whose value now is to be found) 
on demand and cumulative emissions. The demand
process is translated into an emissions process via the bid stack, which
allows us to deduce which generators are active at any point in time. As
noted above, the bid stack both influences and is influenced by the allowance
price. This leads naturally to a formulation of the allowance price as the
backward part of a forward-backward stochastic differential equation
(FBSDE). To solve the problem numerically, we derive a semilinear partial
differential equation (PDE) for the allowance price as a function of demand
and cumulative emissions, and we give a formal asymptotic description of the
solution behavior near the end of a compliance period, highlighting the way
in which the nonlinearity in the governing PDE --- which is a consequence of
the feedback from allowance prices to the behavior of energy producers ---
leads to a nonzero probability that the total cumulative emissions hit the
cap exactly (cf. \cite{rCarmona2010g,rCarmona2011b}). In a sense, the market functions so as to produce the maximum
emissions possible without incurring the penalty and this is an important
practical consequence of our analysis.
We extend our model to emission markets with multiple
compliance periods and analyze the impact different intertemporal connecting
mechanisms such as borrowing, banking, and withdrawal have on the allowance
price.  The last section is devoted to the pricing of European derivatives
written on the allowance certificate. Throughout the analysis we focus on the
trading of AAUs only. The joining of multiple markets using CERs and ERUs in
the present setting is left to future research and is addressed from a
different point of view in, for example, \cite{rCarmona2011a}.

\section{From electricity markets to carbon emissions}
In this section we develop our approach to modeling the interaction between electricity and emission markets. We introduce the random factors and the
key parameters that are later shown to drive the price formation of allowance certificates. An important part is played by the merit order---the
rule by which available resources with the lowest marginal costs of production are called upon first to supply electricity. We introduce the
electricity bid stack, which is modeled as a continuous map from the supply of electricity to its marginal price, and analogously define the emissions
stack as a continuous map to the marginal emissions caused by the production of the last unit. Using an equilibrium assumption, we relate supply to
demand; we show how this allows us to deduce which technologies are used to meet demand at any point in time and the total market emissions rate this
production schedule implies. Finally, we illustrate the impact the introduction of a cost of carbon has on the bid and emissions stacks. In the context
of an emissions trading scheme the merit order assumption very naturally leads to load shifting, the reallocation of energy production from
emission-intensive to pollution friendly resources.

\subsection{Market set-up}
We consider a finite time interval $[0,T]$, which initially corresponds to one compliance period; later we will consider multiperiod markets. We
denote by $(\Omega,\mathcal{F},(\mathcal{F}_t)_{t\in[0,T]}, \mathbb{P})$ a filtered probability space satisfying all the usual assumptions, where
$(\mathcal{F}_t)_{t\in[0,T]}$ is generated by a standard Brownian motion $(W_t)_{t\in[0,T]}$, the only source of randomness in the market. In order to
simplify the notation we omit the subscript that restricts a stochastic process to the time interval $[0,T]$ from now on. We deviate from this
habit only in section \ref{str:allowance_pricing_multiple_periods}, where it becomes important as part of a multiperiod setting, and in section
\ref{str:option_pricing}, where we discuss the pricing of derivatives.

Agents in our market demand a good, the production of which causes emissions; as discussed above, we take this good to be electricity. Firms can
produce electricity using different technologies that vary in their costs of production and their emissions intensity. The market is subject to an
emissions trading scheme, as follows. Each registered firm receives an initial allocation of allowances, which can be used to offset its cumulative
emissions at the end of the compliance period. If a firm is unable to submit a sufficient number of certificates, its excess emissions are subject to
the payment of a monetary penalty. Allowances are represented by printed certificates. Because their cost of carry is negligible, we consider them to
be liquidly traded financial products in which long and short positions can be taken. Consequently, if a firm believes its initial allocation to be
incorrect, it can buy or sell allowances as needed. This leads to a liquid market and the formation of a price at which allowances are traded.

Analogous to the idea of a representative agent, we ignore the aggregation problem and instead take the point of view of the whole market. Our goal
then becomes to determine the arbitrage-free price of emission permits as a function of the aggregate forces that act in the market. As will be shown
in section \ref{str:allowance_pricing}, this price directly and crucially depends on the accumulated emissions during the compliance period and on the
aggregate demand for electricity.

The actions of consumers in the market result in an exogenously given $\mathcal{F}_t$-adapted demand process $(D_t)$. Firms respond to this demand by
generating electricity. In particular, at any time $0\leq t \leq T$, the aggregate of all firms supplies an amount $(\xi_t)$ of electricity. We assume
that the market uses only currently available information to decide on its production level and that this level is always nonnegative and below a
constant maximum production capacity $\xi_{\max}\geq 0$. Therefore, $\xi_t$ is $\mathcal{F}_t$-adapted, and
\begin{equation*}
 0\leq \xi_t \leq \xi_{\max} \quad \text{for } 0 \leq t \leq T.
\end{equation*}
Moreover, we assume that there are always sufficient resources in the market to meet demand so that
\begin{equation*}
 0 \leq D_t \leq \xi_{\max} \quad \text{for } 0 \leq t \leq T.
\end{equation*}
The demand process is assumed to be perfectly inelastic, as is frequently justifiable in electricity markets (cf. \cite{rCarmona2010f, mCoulon2009a}),
and demand and supply are related by a Walrasian equilibrium assumption (cf. \cite{lWalras1952}). This concept is realized by the market
administrator, who ensures that aggregate demand for and aggregate supply of energy are matched on a daily basis, namely, that
\begin{equation}\label{eq:walrasian_Equilibrium}
 D_t = \xi_t \quad \text{for } 0 \leq t \leq T.
\end{equation}
Typically, spot data for demand and supply is quoted in megawatts. For example, a demand of 60MW for one hour is equivalent to 60MWh.

The production of electricity causes CO$_2$ emissions in a way that we describe more precisely in sections \ref{str:stacks} and
\ref{str:load_shifting}. The total (cumulative) emissions during the time interval $[0,t]$ are described by the process $(E_t)$, which is measured in
tonnes of CO$_2$. Moreover, since emission intensive production resources are 
finite and demand is bounded, $(E_t)$ is also bounded; i.e., 
\begin{equation*}
 0\leq E_t \leq E_{\max} \quad \text{for } 0 \leq t \leq T.
\end{equation*}

The regulator decides on an acceptable maximum level of cumulative emissions
during the compliance period (the cap) and issues a corresponding number of
allowance certificates, $0\leq E_{\text{cap}} \leq E_{\max}$, measured in
tonnes of CO$_2$. At the end of the compliance period, cumulative emissions
in the market are offset against the initial allocation of
allowances. Certificates that are not used for this purpose expire worthless
in the case of the single-period setup, whereas unaccounted emissions are
subject to a monetary penalty payment at a rate $\Pi\geq 0$. Thus, an amount
$(E_T-E_{\text{cap}})^+$ of emissions is penalized.

The allowance certificates constitute traded assets in the market. Their value is represented by the process $(A_t)$. We shall also consider options
written on the certificate and assume the existence of a riskless money market account with constant risk-free rate $r\geq 0$.

\subsection{The bid and emissions stacks}\label{str:stacks}
We turn to the modeling of the cumulative emissions. We begin with the business-as-usual market and analyze the impact of an emissions trading scheme
in the next subsection.

Key to our analysis is the following assumption, which summarizes the actions of the central market administrator as introduced above.

\begin{assumption}\label{as:merit_order}\rm The market administrator ensures that resources are used according to the \textit{merit order}. This means
that the cheapest production technologies are called upon first to satisfy a given demand, and hence electricity is supplied at the lowest possible
price.
\end{assumption}

As explained in section \ref{str:intro_el}, bid levels are mostly determined by variable costs. Therefore, these costs play an integral part in
determining the merit order arrangement in Assumption \ref{as:merit_order}. The resulting increasing map from market supply of electricity to marginal
price forms the bid stack. As explained in the introduction, the bid stack is, strictly speaking, an increasing simple function. In practice, however,
it consists of sufficiently many steps to be approximated by a smooth function. This leads us to the following definition.

\begin{deff}\label{def:BAU_stack}The {\rm{business-as-usual bid
      stack}} is given by the continuous function
 \begin{equation*}
  b^{\text{BAU}}(\xi) : [0,\xi_{\max}] \mapsto [0,\infty),
 \end{equation*}
 where $b^{\text{BAU}}(\cdot)\in C^1(0,\xi_{\max})$ and $\Id b^{\text{BAU}}/ \Id \xi >0$.
\end{deff}

Here and throughout the rest of the paper, the variable $\xi$ represents the
supply of electricity (measured in MW). Correspondingly,
$b^{\text{BAU}}(\xi)$ denotes the bid level of the marginal production unit
(measured in \euro\ per MWh).

We note immediately that in reality business-as-usual bid levels are
stochastic. Most importantly, fuel prices, which are key drivers of variable
costs, fluctuate continuously. In principle the model that we propose can be
extended to include stochastic fuel prices as part of the variable costs that
determine firms' bids. The business-as-usual bid stack $b^{\text{BAU}}$ would
then become a function of additional independent variables (the prices of the
fuels used in the production process), and the dimensionality of the
allowance pricing problem \eqref{eq:allowance_FBSDE} would increase. Such an
extension should be considered when one is interested in pricing contracts
such as, for example, clean spread options, which explicitly feature the
prices of electricity, fuels, and emissions in their payoff. In this case the
subtle dependence of electricity spot prices on fuel prices becomes
important. Since we are predominantly interested in the price formation of
allowance certificates, we only mention the possibility of this extension and
leave its investigation to future research.\footnote{Since the
  original publication of this article in \cite{sHowison2012} research
  in this direction has been undertaken
  (cf. \cite{rCarmona2011}).}

In the current paper we are interested
only in the relative position of the different technologies in the bid
stack. Fluctuations in fuel prices become important only if they induce merit
order changes. From historic data observations this is relevant only in the
long run, and we prefer not to consider it for now. Hence we model the
business-as-usual bid stack as a deterministic function (cf.\
\cite{mBarlow2002}), allowing us to focus exclusively on the impact of
emissions trading on variable costs and the merit order in section
\ref{str:load_shifting}.

\begin{remark}\label{rem:BAU_merit_order}\rm As pointed out in the
  introduction, emission-intensive technologies tend to be cheaper than
  environmentally friendly ones as a means to produce electricity. Therefore,
  we find that bids associated with a small level of electricity supply stem
  mostly from emission-intensive generators, while bids at the right end of
  the interval $[0,\xi_{\max}]$ stem mostly from environmentally friendly
  ones (as remarked earlier, exceptions to this rule are nuclear plants,
  which do not cause any CO$_2$ emissions and are generally placed at the
  very left end in the bid stack). In between exists a spectrum in which a
  mixture of technologies contributes to bids. This assumption has been
  confirmed (cf.\ \cite{mCoulon2009a}) by analyzing the correlation between
  production costs and bid levels.
\end{remark}

Analogous to the bid stack, we construct an emissions stack by creating a map from the supply of electricity to the marginal emissions associated with
the supply of the last unit.

\begin{deff}\label{def:ems_stack}The {\rm{marginal emissions stack}}
  is given by the continuous function
 \begin{equation*}
 e(\xi) : [0,\xi_{\max}] \mapsto (0,\infty),
 \end{equation*}
where $e(\cdot)\in C^1(0,\xi_{\max})$.
\end{deff}

With the above definition, $e(\xi)$ associates with a specific supply of electricity $\xi$ the emissions rate of the marginal unit (measured in tonnes
of CO$_2$ per MWh).

\begin{prop}The business-as-usual \textit{market emissions rate} $\mu_E^{\text{BAU}}$ is given by
 \begin{equation*}
 \mu_E^{\text{BAU}}(D) := \kappa \int_0^{D} e(\xi)\ \Id \xi \quad \text{for } 0\leq D \leq \xi_{\max},
 \end{equation*}
where the scaling constant $\kappa$ is the ratio of the emissions
period $T$ to that of the time unit associated with the marginal
emissions stack $e$ (typically, $T$ is measured in years and $\kappa$ is the 
  number of hours per year).
\end{prop}

\begin{proof}
  The Walrasian equilibrium assumption \eqref{eq:walrasian_Equilibrium} for
  our inelastic model implies that the market produces the exact amount of
  electricity consumers demand and that---under business-as-usual---the
  generation capacity associated with the interval $[0,D]$ is used for this
  purpose. The market emissions rate per hour is then obtained at any time by
  integrating over the marginal emissions stack up to the current level of
  demand. We rescale this rate with $\kappa$
  so that $\mu_E^{\text{BAU}}$ is the market emissions rate per unit
  of $T$.
\end{proof}


\subsection{Load shifting: A short-term abatement measure}\label{str:load_shifting}
We now analyze the effects of emissions trading on the business-as-usual economy introduced above. As explained in the introduction, emissions trading
puts a price on carbon and thereby increases the production costs of firms. In particular, it makes it more expensive for firms that rely on
emission-intensive technologies to produce. For each unit of CO$_2$ that these firms emit in excess of their initial allocation, they must buy an
allowance contract in order to avoid penalization; the cost of carbon is a real cost. Alternatively, if a firm owns more allowances than it requires,
it can sell spare ones in the market. In this case, the cost of carbon represents an opportunity cost.

We ignore the possibility that firms might invest in long-term abatement projects and focus only on the direct impact on the bid stack. We assume
that, in order to maintain their profit margin, firms pass the emissions-related increase in production costs on to consumers. Because the cost of
carbon is represented by the price of an allowance certificate, the business-as-usual bids of each firm increase by an amount equal to the allowance
price multiplied by the marginal emissions rate of that firm. On an aggregate level this means that, for a given allowance price $A$ the bid stack now
becomes the function $g$, where
\begin{equation}\label{eq:bid_stack_emissions_trading}
 g(A,\xi):=b^{\text{BAU}}(\xi) + A e(\xi) \quad \text{for } 0\leq A < \infty,\ 0 \leq \xi \leq \xi_{\max}.
\end{equation}
For $A=0$, \eqref{eq:bid_stack_emissions_trading} is equivalent to the business-as-usual bid stack. For positive certificate prices emissions trading
may cause the mapping $\xi \mapsto g(\cdot,\xi)$ to lose its monotonicity. In particular, we observe that bids associated with large marginal emission
rates become relatively more expensive, as the cost of carbon makes it relatively more costly for firms relying on dirty fuels, such as coal, to
produce.

By the merit order assumption the market administrator calls upon generators in increasing order of their bid levels. We define the set of active
generation units at a given allowance and electricity price $P$ by
\begin{equation}\label{eq:active_generation_units}
 S(A,P) := \left\{ \xi \in [0,\xi_{\max}] : g(A,\xi) \leq P\right\} \quad \text{for } 0\leq A < \infty,\ 0\leq P < \infty.
\end{equation}
By the definition of a sublevel set, $P\mapsto\lambda(S(\cdot,P))$, where $\lambda$ denotes the Lebesgue measure, is strictly increasing; under the
following assumption, it is also continuous and therefore invertible.

\begin{assumption}\label{as:b_measure_zero}\begin{equation*}
\lambda\left(\left\{\xi\in (0,\xi_{\max}) : \frac{\partial b^{\text{BAU}}}{\partial \xi}(\xi) + A\frac{\partial e}{\partial \xi}(\xi) =
0\right\}\right) = 0 \quad \text{for } 0\leq A <\infty.
\end{equation*}
\end{assumption}

Using \eqref{eq:active_generation_units}, for observed values of the allowance price, the market bid stack $b$ is now defined by
\begin{equation*}
 b(A,\xi) := \lambda(S(A,\cdot))^{-1}(\xi) \quad \text{for } 0\leq A < \infty,\ 0\leq \xi \leq \xi_{\max}.
\end{equation*}
This immediately yields the market price of electricity $P$, which is given by
\begin{equation*}
P:=b(A,D) \quad \text{for } 0\leq A < \infty,\ 0\leq D \leq \xi_{\max}.
\end{equation*}

Whereas under business-as-usual demand $D$ is met using the generation capacity $[0,D]$ (considered a subset of the domain of the emissions stack
$e$), emissions trading may shift this interval further to the right, or, depending on the shape of the marginal emissions stack, split it up into
multiple sets with combined Lebesgue measure $D$---an effect we refer to as load shifting. We make the impact of load shifting on the market emissions rate $\mu_E$ precise in the next
proposition.

\begin{prop}\label{prop:inst_emissions}In the presence of cap-and-trade and given an allowance price $A$ and demand level $D$, the market emissions
rate $\mu_E$ is given by
\begin{equation}\label{eq:inst_emissions}
 \mu_E(A,D) = \kappa\int_{S_p(A,D)}e(\xi)\ \Id \xi \quad \text{for } 0\leq A < \infty,\ 0\leq D \leq \xi_{\max},
\end{equation}
where $S_p(A,D):= S(A,b(A,D))$.
\end{prop}

\begin{proof}
The proof is immediate from the discussion above.\qquad\end{proof}

We note that the business-as-usual market emissions rate is of course a special case of \eqref{eq:inst_emissions}, which is obtained by setting $A=0$,
in which case $S_p(0,D)=[0,D]$.

\begin{remark}\rm As described earlier, in reality the bid and marginal emissions stack are step functions, whose finitely many constant values
  correspond to firms' bids and their corresponding marginal emissions. To
  model the impact of a positive allowance price on the bid stack in this
  case, one would add the cost of carbon to bids as usual, and then the
  resulting step function is \textit{rearranged} in increasing order. Because
  of the discrete nature of the problem, the rearrangement induces a
  permutation $\nu$ on the bids, which is then applied to the marginal
  emissions stack.  Instantaneous emissions are now obtained by integrating
  the rearranged emissions stack over the closed interval $[0,D]$. We prefer
  to work with the continuous limit of the bid and marginal emissions
  stack. In this case the permutation $\nu$ cannot be defined explicitly, and
  we identify active firms with the set $S_p$.
\end{remark}

In the following lemma we prove some technical properties of $\mu_E$, which show that the model we propose for the market emissions rate makes
intuitive sense and leads to a suitably regular function.

\begin{lem}The market emissions rate $\mu_E$ satisfies the following.
\begin{enumerate}[\rm(L.1)]
\item The map $D \mapsto \mu_E(\cdot,D)$ is
\begin{enumerate}[\rm(i)]
\item strictly increasing and
\item Lipschitz continuous.
\end{enumerate}
\item The map $A \mapsto \mu_E(A,\cdot)$ is
\begin{enumerate}[\rm(i)]
\item nonincreasing and
\item Lipschitz continuous.
\end{enumerate}
\item $\mu_E$ is bounded.
\end{enumerate}
\end{lem}

\begin{enumerate}[(L.1)]
 \item
 \begin{enumerate}[(i)]
 \item By Assumption \ref{as:b_measure_zero} and the definition of sublevel set, for $0\leq D_1 < D_2\leq \xi_{\max}$, $S_p(\cdot,D_1)\subset
S_p(\cdot,D_2)$. Since $e(\xi) > 0$ on $[0,\xi_{\max}]$ the result follows.
 \item For $0\leq D_1 < D_2 \leq\xi_{\max}$ and with the definition $\Delta^D S_p(D_2,D_1):=S_p(\cdot,D_2)\setminus S_p(\cdot,D_1)$,
 \begin{align*}
  \mu_E(\cdot,D_2) - \mu_E(\cdot,D_1) &= \kappa\int_{\Delta^D S_p(D_2,D_1)} e(\xi)\ \Id \xi\\
  & \leq \lambda\left(\Delta^D S_p(D_2,D_1)\right) \kappa\max_\xi e(\xi)\\
  & = (D_2-D_1)\kappa\max_\xi e(\xi).
 \end{align*}
 The case $D_2<D_1$ is treated similarly.
 \end{enumerate}
 \item
 \begin{enumerate}[(i)]
 \item \label{lemproof:decreasing_A}For $0\leq A_1 < A_2 < \infty$ and with the definition $\Delta^A S_p(A_1,A_2):=S_p(A_1,\cdot)\setminus
S_p(A_2,\cdot)$,
 \begin{equation*}
  \mu_E(A_1,\cdot) - \mu_E(A_2,\cdot) = \kappa\int_{\Delta^A S_p(A_1,A_2)} e(\xi)\ \Id \xi - \kappa\int_{\Delta^A S_p(A_2,A_1)} e(\xi)\ \Id \xi.
 \end{equation*}
 Since $\lambda(\Delta^A S_p(A_1,A_2)) = \lambda(\Delta^A S_p(A_2,A_1))$, the result follows from the observation that, for a given $0\leq D \leq
\xi_{\max}$, $e(\xi) = (g(A_2,\xi)-g(A_1,\xi))(A_2-A_1)^{-1} > (b(A_2,D)-b(A_1,D))(A_2-A_1)^{-1}$ on $\Delta^A S_p(A_1,A_2)$ and that $e(\xi) =
(g(A_2,\xi)-g(A_1,\xi))(A_2-A_1)^{-1} \leq (b(A_2,D)-b(A_1,D))(A_2-A_1)^{-1}$ on $\Delta^A S_p(A_2,A_1)$.
 \item From above we know that
 \begin{equation*}
  \mu_E(A_1,\cdot) - \mu_E(A_2,\cdot) \leq C_1 \lambda(\Delta^A S_p(A_1,A_2))
 \end{equation*}
 for some constant $C_1\geq 0$. It is also clear that $\Delta^A S_p(A_1,A_2)$ (and similarly $\Delta^A S_p(A_2,A_1)$) can be written as the union of a
finite number of intervals. As $A_1$ increases to $A_2$, there are three possibilities: (a) existing intervals grow or shrink; (b) new intervals
appear, or existing ones disappear (by Assumption \ref{as:b_measure_zero} this always happens at a point); and (c) the intervals remain unchanged.
Differentiating the level curves $g(A,\xi) = b(A,D)$ with respect to $\xi$, for a given level of demand, we find that
 \begin{equation*}
  \frac{\Id \xi}{\Id A}= - \left(\frac{\partial g}{\partial A} - \frac{\partial b}{\partial A}\right)\bigg/\frac{\partial g}{\partial \xi}.
 \end{equation*}
 By Assumption \ref{as:b_measure_zero} the right-hand side is bounded by a constant,\footnote{Throughout this proof we allow $C_2$ to change from
occurrence to occurrence.} say, $C_2\geq 0$. Therefore, as $A$ changes, in each case (a)--(c), the endpoints of the intervals defining $\Delta^A
S_p(A_2,A_1)$ do not move faster than $C_2(A_2-A_1)$. Therefore, also $\lambda(\Delta^A S_p(A_1,A_2))\leq C_2(A_2-A_1)$. The case $A_1 > A_2$ is
treated similarly, and the result follows.
 \end{enumerate}
 \item Boundedness of $\mu_E$ follows from the boundedness of $e$ and the fact that $S_p(A,D)\subseteq [0,\xi_{\max}]$ for all $A\geq 0$ and $0 \leq
D \leq \xi_{\max}$.\qquad\endproof
 \end{enumerate}

From the definition of instantaneous emissions we derive cumulative emissions by integrating over \eqref{eq:inst_emissions}, up to the current time
$t$.

\begin{figure}[htbp]
  \centering
  \subfloat[Bid stacks $b^{\text{BAU}}$ and $g$.]{\label{fig:ex_bids_rearr_A100}\includegraphics[width=0.72\textwidth]{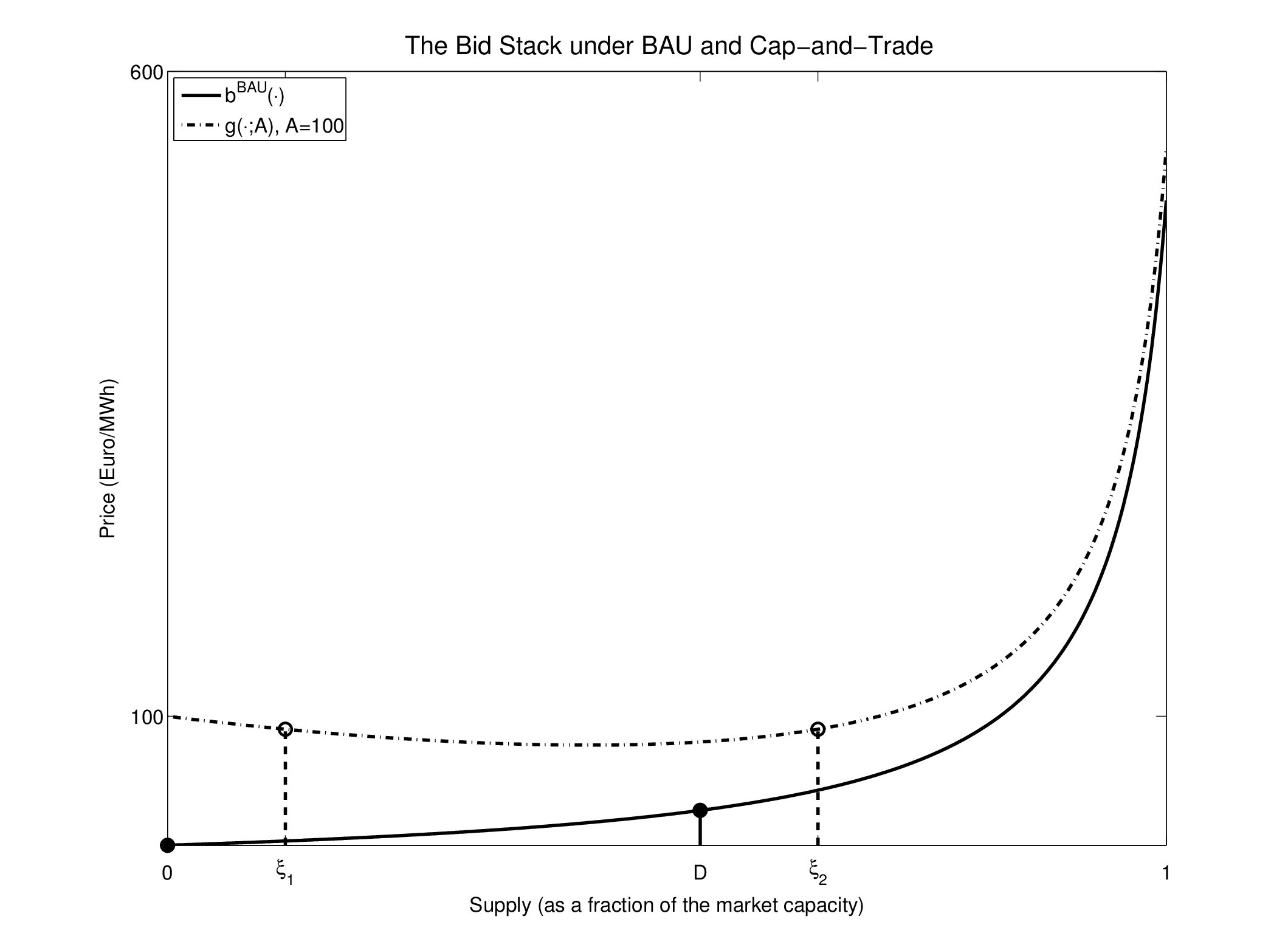}}\\
  \subfloat[Emissions stack $e$.]{\label{fig:ex_ems_rearr_A100}\includegraphics[width=0.72\textwidth]{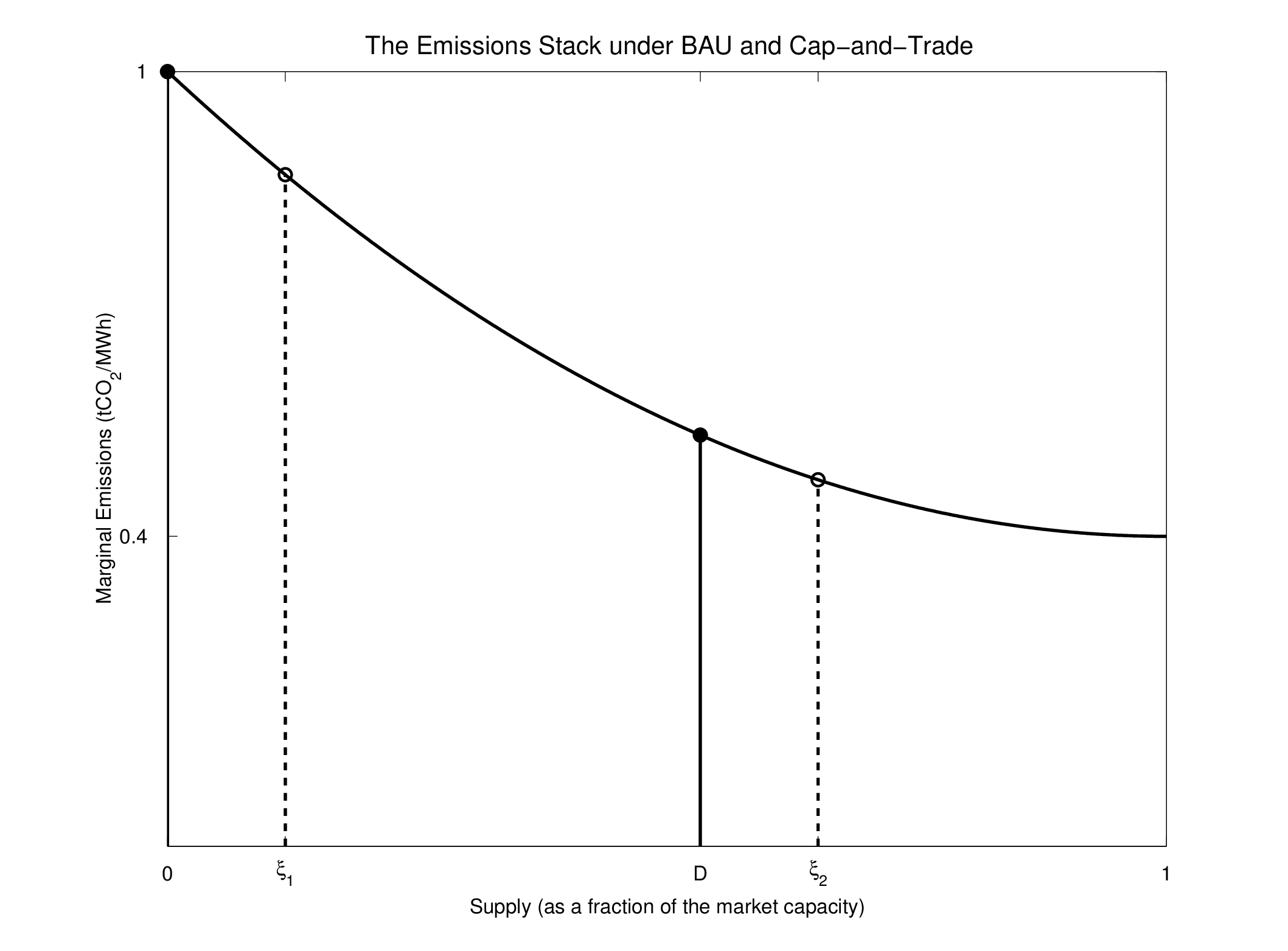}}
  \caption{Under business-as-usual conditions, the bid stack $b^{\text{BAU}}$ implies that resources associated with the interval $[0,D]$ are used to
meet demand. Therefore, instantaneous emissions are obtained by integrating over the emissions stack from $0$ to $D$. Under the influence of a
cap-and-trade scheme, the function $b$ leads to resources being shifted to the interval $[\xi_1,\xi_2]$. Instantaneous emissions are now given by the
(smaller) integral over the emissions stack from $\xi_1$ to $\xi_2$.}
 \label{fig:ex_stacks}
\end{figure}

Figure \ref{fig:ex_stacks} illustrates the effect of load shifting and the resulting reduction in the market emissions rate under the assumption that
under BAU dirtier production technologies are placed further to the left in the bid stack than cleaner ones (see Remark \ref{rem:BAU_merit_order}).

\section{Risk-neutral pricing of allowance
  certificates}\label{str:allowance_pricing}
In this section we address the problem of determining the arbitrage-free price of an allowance certificate given the current demand for electricity
and the cumulative emissions to date in the economy. (Recall that the
notion of an arbitrage-free price rules out the possibility of making
a profit starting with no initial investment.) We do this initially in the setting of an emissions market with one compliance period;
subsequently we generalize the model to deal with markets that consist of multiple consecutive compliance periods and examine the impact that
connecting mechanisms, namely, banking, borrowing, and withdrawal,
have on the certificate price. One of the groundbreaking results in
the field of mathematical finance was the realization that the absence
of arbitrage is in fact equivalent to the existence of a very
particular probability measure, say $\mathbb{Q}$, on
$(\Omega,\mathcal{F})$ (cf. \cite{jHarrison1981,fDelbaen1994}). This measure is equivalent to $\mathbb{P}$,
meaning that $\mathbb{P}(N) = 0$ if and only if $\mathbb{Q}(N) = 0$,
and has the property that the discounted prices of all tradable assets
(the allowance certificates in our case) are martingales under $\mathbb{Q}$. This motivates our next assumption.

\begin{assumption}\label{as:traded_assets}\rm There exists an equivalent martingale measure $\mathbb{Q}\sim\mathbb{P}$, under which, for $0 \leq t
\leq T$, the discounted price of any tradable asset is a martingale. We refer to $\mathbb{Q}$ as the \textit{risk-neutral measure}.
\end{assumption}

We begin by making some additional assumptions about the demand and
cumulative emissions processes $(D_t)$ and $(E_t)$. We assume that at time
$t=0$ demand for electricity is known. Thereafter, it evolves according to an
It\^{o} diffusion; i.e., for $0\leq t \leq T$, under the measure
$\mathbb{Q}$, demand for electricity is given by the stochastic process
\begin{equation}\label{eq:demand_process}
 \Id D_t = \mu_D(D_t) \Id t + \sigma_D(D_t) \Id \tilde{W}_t, \qquad D_0=d \in (0,\xi_{\max}),
\end{equation}
where $(\tilde{W}_t)$ is $\mathcal{F}_t$-adapted and a $\mathbb{Q}$-Brownian
motion (we postpone the discussion of the relevance of the regularity of the
coefficients to section \ref{str:allowance_pricing_one_cp}). The assumption
that demand is perfectly inelastic is reflected in the fact that both
coefficients are functions of demand only. Note that 
if there were a feedback from
price to demand in the model then additional nonlinearities to those we see
below would arise. Note also that in practice demand for electricity
exhibits seasonal periodicity, an attribute that would cause $\mu_D$ to
depend on time explicitly. For simplicity we choose to ignore this feature.

Cumulative emissions are measured from the beginning of the compliance period when time $t=0$ so that $E_0=0$. Subsequently, they are determined by
integrating over the market emissions rate $\mu_E$ derived in
Proposition \ref{prop:inst_emissions}. Consequently, the cumulative
emissions process is represented by an absolutely continuous process; i.e., for $0\leq t \leq T$,
\begin{equation}\label{eq:cumulative_emissions_process}
 \Id E_t = \mu_E(A_t,D_t) \Id t, \quad E_0=0.
\end{equation}
Note that with this definition the process $(E_t)$ is nondecreasing, which makes intuitive sense considering that it represents a cumulative
quantity.

\subsection{One compliance period}\label{str:allowance_pricing_one_cp}
To formulate the pricing model, it remains to characterize the allowance certificate price process $(A_t)$. This is different from the specification
of $(D_t)$ and $(E_t)$, because its value at time $t=0$ is unknown. An arbitrage argument, however, allows us to determine its value at the end of the
compliance period. The event of noncompliance is $\{E_T\geq E_{\text{cap}}\}$; then the value of the allowance certificate at time $t=T$ is given by
the terminal condition
\begin{equation}\label{eq:allowance_terminal_cnd}
A_T =
\begin{cases}
 0 &\text{for } 0\leq E_T < E_{\text{cap}},\\
 \Pi &\text{for } E_{\text{cap}} \leq E_T \leq E_{\max}.
\end{cases}
\end{equation}

From Assumption \ref{as:traded_assets} we know that the discounted allowance price is a martingale under the measure $\mathbb{Q}$. Therefore, the
allowance price is given as the discounted conditional expectation of its terminal condition under this measure; i.e.,
\begin{equation}\label{eq:allowance_risk_neutral_expectation}
 A_t=e^{-r(T-t)}\Pi\,\E^\mathbb{Q}\left[\left.\mathbb{I}_{[E_{\text{cap}},\infty)}(E_T)\right|\mathcal{F}_t\right] \quad \text{for } 0\leq t \leq T,
\end{equation}
which shows that the allowance price process $(A_t)$ takes values in $[0,\Pi]$ only.

\begin{prop}\label{prop:allowance_price_one_period}For $0 \leq t \leq T$, the price of an allowance certificate $(A_t)$ in a market with one
compliance period is described by the following FBSDE:
\begin{equation}\label{eq:allowance_FBSDE}
\left\{
 \begin{aligned}
 \Id D_t &= \mu_D(D_t) \Id t + \sigma_D(D_t) \Id \tilde{W}_t,	&& D_0=d\in (0,\xi_{\max}),\\
 \Id E_t &= \mu_E(A_t,D_t) \Id t,				&& E_0=0,\\
 \Id A_t &= rA_t \Id t + e^{rt}Z_t \Id \tilde{W}_t,		&& A_T = \Pi\,\mathbb{I}_{[E_{\text{cap}},\infty)}(E_T).
 \end{aligned}
\right.
\end{equation}
\end{prop}

\begin{proof}
Because the filtration $(\mathcal{F}_t)$ is natural, it is a consequence of the Martingale Representation Theorem (cf. \cite{iKaratzas1999}) that the
discounted allowance price can be represented as an It\^{o} integral with respect to the Brownian motion $(\tilde{W}_t)$. It follows that
\begin{equation}\label{eq:FBSDE_derivation_1}
 \Id \left(e^{-rt}A_t\right) = Z_t \Id \tilde{W}_t \quad \text{for } 0\leq t \leq T
\end{equation}
for some $\mathcal{F}_t$-adapted process $(Z_t)$.

Combining the processes \eqref{eq:demand_process} and \eqref{eq:cumulative_emissions_process} for demand and cumulative emissions and
\eqref{eq:FBSDE_derivation_1} together with the terminal condition \eqref{eq:allowance_terminal_cnd}, the pricing problem becomes that described by
\eqref{eq:allowance_FBSDE}\mbox{.\qquad}\end{proof}

\begin{remark}\rm The existence and uniqueness of a solution to the FBSDE \eqref{eq:allowance_FBSDE} is a delicate question. The nonstandardness of this kind of equation arises from the degeneracy of one of its forward components
(the emissions process $(E_t)$ in our case) combined with the singularity of the terminal condition. Together, these features conspire to cause the
random variable $E_T$ to develop a point mass at the cap $E_{\text{cap}}$, as shown in \cite{rCarmona2011b}. In the same paper it is also shown that
under the assumption that $\mu_D$ and $\sigma_D$ are Lipschitz continuous and exhibit at most linear growth and that $\mu_E$ is Lipschitz continuous
and strictly decreasing in $A$, a unique solution to \eqref{eq:allowance_FBSDE} exists satisfying the initial conditions $D_0=d$, $E_0=0$ and
the relaxed terminal condition
 \begin{equation*}
 \Pi\,\mathbb{I}_{(E_{\text{cap}},\infty)}(E_T)\leq A_T \leq \Pi\,\mathbb{I}_{[E_{\text{cap}},\infty)}(E_T).
 \end{equation*}
Moreover, it is shown that there exists a continuous function $\alpha$ such that $A_t = \alpha(t,D_t,E_t)$ for $0\leq t < T$. Under considerably more
restrictive conditions on the coefficients but preserving the distinctive features of the problem (degeneracy of the forward component and a
singularity in the terminal condition), the value function $\alpha$ is
actually smooth (cf.~\cite{rCarmona2010g}). Since the original
acceptance of this paper for publication (cf. \cite{sHowison2012}), new results have been obtained
which affirmatively answer the question of existence and uniqueness of a solution to the FBSDE (\ref{eq:allowance_FBSDE}) under 
weaker conditions on the regularity of the coefficients $\mu_D$ and
$\sigma_D$ than required in \cite{rCarmona2011b} and
\cite{rCarmona2010g}. In fact it is sufficient for $\mu_D$ and
$\sigma_D$ to exhibit sufficient regularity to guarantee that the
stochastic differential equation for $(D_t)$ has a strong solution. We refer the interested reader to the 
thesis \cite{newref} for the precise statement and the proof of the theorem.
\end{remark}

Based on the previous remark, we assume that in our
Markovian setting there exists a function
$\alpha(t,D,E):[0,T]\times[0,\xi_{\max}]\times [0,E_{\max}]\mapsto [0,\Pi]$,
such that $A_t=\alpha(t,D_t,E_t)$, for $0\leq t < T$, suitably regular on
$[0,T)$ to be a classical solution to the PDE
\begin{align}\label{eq:allowance_PDE}
 \mathcal{N}\alpha
&=0 && \text{on 
$U$}, \;  0\leq t<T, 
\nonumber\\
 \alpha &= \Pi\,\mathbb{I}_{[E_{\text{cap}},\infty)}(E) && \text{on $U$}, \;t=T,
\end{align}
where $U:=(0,\xi_{\max})\times (0,E_{\max})$ and
\begin{equation*}
 \mathcal{N}\cdot:= \frac{\partial \cdot}{\partial t} 
+ \frac{1}{2}\sigma_D^2(D) \frac{\partial^2 \cdot}{\partial D^2} 
+  \mu_D(D) \frac{\partial \cdot}{\partial D} +
\mu_E(\cdot,D) \frac{\partial \cdot}{\partial E} - r\cdot.
\end{equation*}
Notice that $\mu_E$ depends on $\alpha$; hence the PDE is semilinear (and, in
the absence of a second $E$-derivative, degenerate parabolic). 
In addition to the terminal condition, suitable boundary
conditions have to be supplied. These depend on the specification of the
coefficients of the PDE, and we postpone the issue to section
\ref{str:num_analysis}, where we discuss the numerical solution of the
problem.

\begin{remark}\rm The intuition behind \eqref{eq:allowance_PDE} is simple. We simply assume that, under $\mathbb{Q}$, $A_t$, being a traded asset, has
a drift equal to the risk-neutral rate (cf.~the last equation of \eqref{eq:allowance_FBSDE}). Then we apply It\^{o}'s formula to
$A_t=\alpha(t,D_t,E_t)$ using the first two equations of \eqref{eq:allowance_FBSDE} and take expectations to derive \eqref{eq:allowance_PDE}. This
procedure is purely formal, because it assumes the existence of a classical solution to the PDE \eqref{eq:allowance_PDE}.
\end{remark}

\subsection{Multiple compliance periods}\label{str:allowance_pricing_multiple_periods}
We now consider the pricing problem in an emission market with two compliance periods. In principle, the results presented in this section can easily
be extended to an arbitrary number of periods. To ease the presentation, however, we choose to present the canonical case. Taking $0=T_0 \leq T_1 \leq
T_2=T$, we consider the two compliance periods $[0,T_1],[T_1,T]$. For simplicity we assume that each period corresponds to one year. As previously,
the $\mathcal{F}_t$-adapted process $(D_t)_{t\in[0,T]}$ represents the aggregate demand for electricity. For $i\in\{1,2\}$ the $\mathcal{F}_t$-adapted
process $(E_t)_{t\in[T_{i-1},T_i]}$ measures the cumulative emissions from the beginning of the $i$th compliance period up to time $t$, and
$(A^i_t)_{t\in[T_{i-1},T_i]}$ represents the price of an allowance certificate for compliance at time $T_i$. Also, we denote by $E^1$ the cumulative
emissions at the end of the first compliance period. Each year, the regulator issues a number $E_{\text{cap}}^i \geq 0$ of allowance certificates and
sets the penalty $\Pi^i\geq 0$.

Demand for electricity is given at time $t=0$ and thereafter evolves continuously throughout the trading period $[0,T]$. Further, we assume that
cumulative emissions are measured from the beginning of each compliance period so that
\begin{equation}\label{eq:cumulative_emissions_multi_period_cond}
 E_{T_{i-1}}:=0, \quad i\in\{1,2\}.
\end{equation}
Finally, we note that each process $(A^i_t)_{t\in[T_{i-1},T_i]}$ corresponds to a different vintage of allowance certificates. If we disregard
mechanisms that connect compliance periods, a certificate issued during the first period is for compliance at time $T_1$ only. However, we now wish to
consider mechanisms that connect compliance periods and permit allowances to be transferred between periods. In this case both vintages of
certificates have a more complex dependence. In particular, the second period allowance price depends on cumulative emissions during not only the
second period but also the previous period, as we describe below. The connecting mechanism is now expressed through the terminal
condition at time $T_1$; for now, we do not determine it explicitly
and denote it by some (possibly singular) function $\phi_1$.

\begin{cor}In a market with two compliance periods, the price $(A_t)_{t\in [T_{i-1},T_i]}$ of an allowance certificate during the $i$th period,
$i\in\{1,2\}$, is described by the following FBSDE:
\begin{equation}\label{eq:allowance_FBSDE_multi_period}
\left\{
\begin{aligned}
\Id D_t &= \mu_D\left(D_t\right) \Id t + \sigma_D \left(D_t\right) \Id \tilde{W}_t, &&D_{T_{i-1}}=d \in (0,\xi_{\max}),\\
\Id E_t &= \mu_E\left(D_t, A^i_t\right) \Id t, &&E_{T_{i-1}} =0,\\
\Id A^i_t &= r A^i_t \Id t + e^{rt}Z^i_t \Id \tilde{W}_t, &&A_{T_i} = \phi_i,
\end{aligned}
\right.
\end{equation}
for some $\mathcal{F}_t$-adapted process $(Z^i_t)_{t\in[T_{i-1},T_i]}$ and where $\phi_1:=\phi_1(E_{T_1})$ and $\phi_2:=\phi_2(E_{T_2};E^1)$,
respectively, denote the terminal condition at the end of the first and second compliance periods.
\end{cor}
\begin{proof}
The proof follows immediately from Proposition \ref{prop:allowance_price_one_period} and the discussion above.\qquad\end{proof}

As in section \ref{str:allowance_pricing_one_cp}, we assume the existence of suitably regular functions
$\alpha_i:[T_{i-1},T_i]\times[0,\xi_{\max}]\times [0,E_{\max}]\mapsto \R_+$ such that $A^i_t=\alpha_i(t,D_t,E_t)$ for $T_{i-1}\leq t < T_i$ and
\begin{align}\label{eq:allowance_PDE_multiple_cp}
 \mathcal{N} \alpha_i &= 0 && \text{on } U,\ T_{i-1} \leq t < T_i\nonumber\\
 \alpha_i &= \phi_i(E) && \text{on }  U,\ t=T_i.
\end{align}

\subsubsection{Banking and withdrawal}\label{str:allowance_pricing_BW}
Banking and withdrawal are two mechanisms that connect compliance periods and are implemented in most emission markets. Both affect the supply of
certificates during the second compliance period. This leads us to introduce $\hat{E}_{\text{cap}}^2$ to denote the aggregate supply of certificates
during the second compliance period. The implementation of banking offers an additional incentive for reducing emissions, since it specifies that
spare allowance certificates, for compliance at the end of the first period, become perfect substitutes for certificates issued during the second
compliance period. This means that in the event of compliance, a number $(E_{\text{cap}}^1-E^1)$ of certificates with price $A^1_{T_1}$ are exchanged
for certificates valid during the next compliance period, with price $A^2_{T_1}$.

This incentive to reduce emissions is strengthened by the withdrawal mechanism, which constitutes additional punishment for firms that exceed their
emission limit. Under this mechanism not only are excess emissions at the end of the first compliance period penalized at the rate $\Pi^1$, but,
moreover, a corresponding number of certificates are withdrawn from the subsequent allocation. Whereas any amount of certificates can be banked, at
most the next period's allocation can be withdrawn from the market. Therefore, in the event of noncompliance, a number $\min(E^1 - E_{\text{cap}}^1,
E_{\text{cap}}^2)$ of certificates with price $A^2_{T_1}$ are subtracted from $E_{\text{cap}}^2$. In the event that the entire allocation of the
second period has been withdrawn and there remain unaccounted-for emissions (at the end of the first period), we specify that these are penalized at
the combined rate of the first period penalty $\Pi^1$ and---to compensate for the lack of certificates that can be withdrawn---an additional
penalty $\bar{\Pi}^1\geq A^2_{T_1}$.

\begin{figure}[htbp]
  \centering
  \subfloat[Banking.]{\label{fig:banking_schematic}\includegraphics[width=0.5\textwidth]{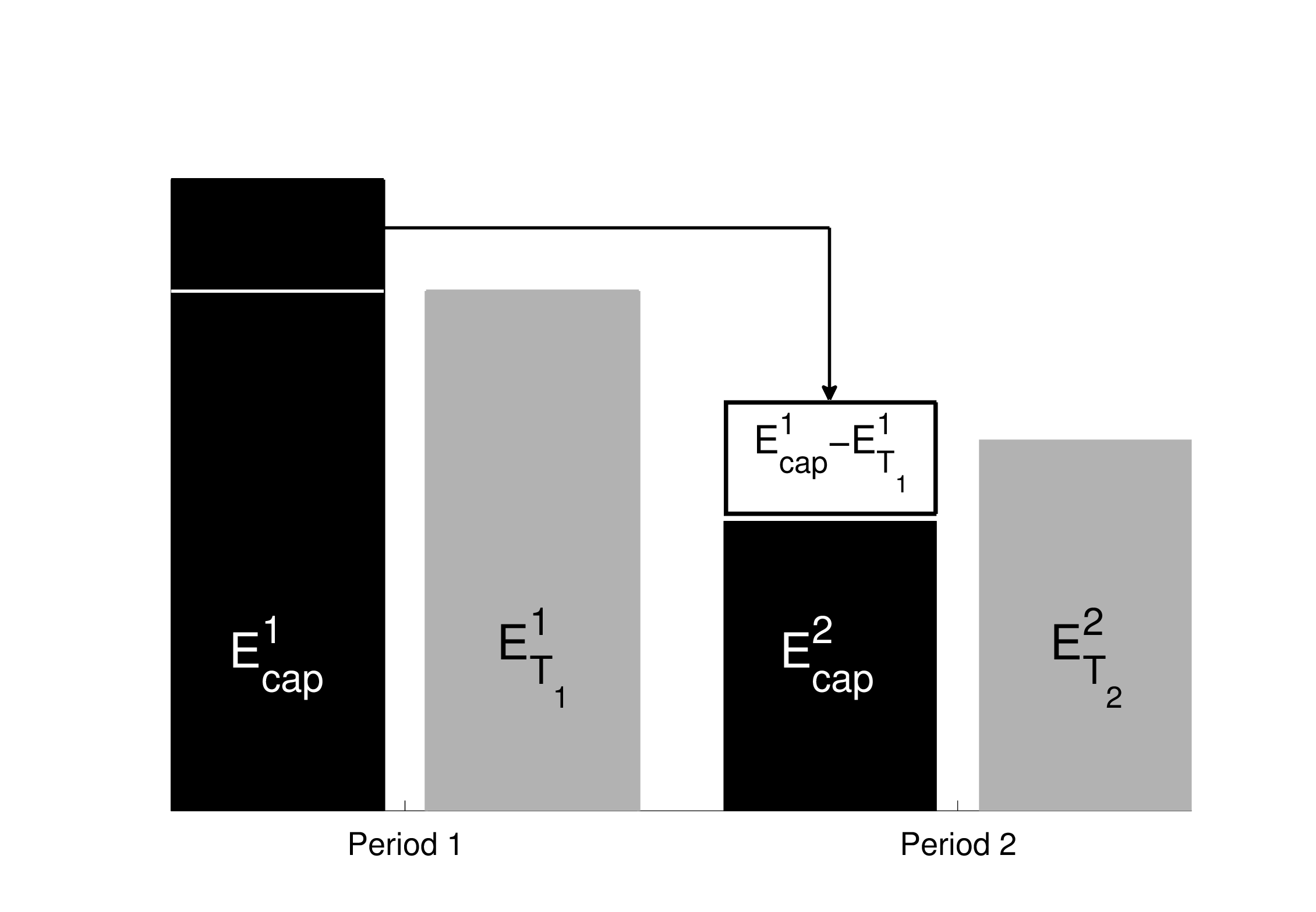}}
  \subfloat[Withdrawal.]{\label{fig:withdrawal_schematic}\includegraphics[width=0.5\textwidth]{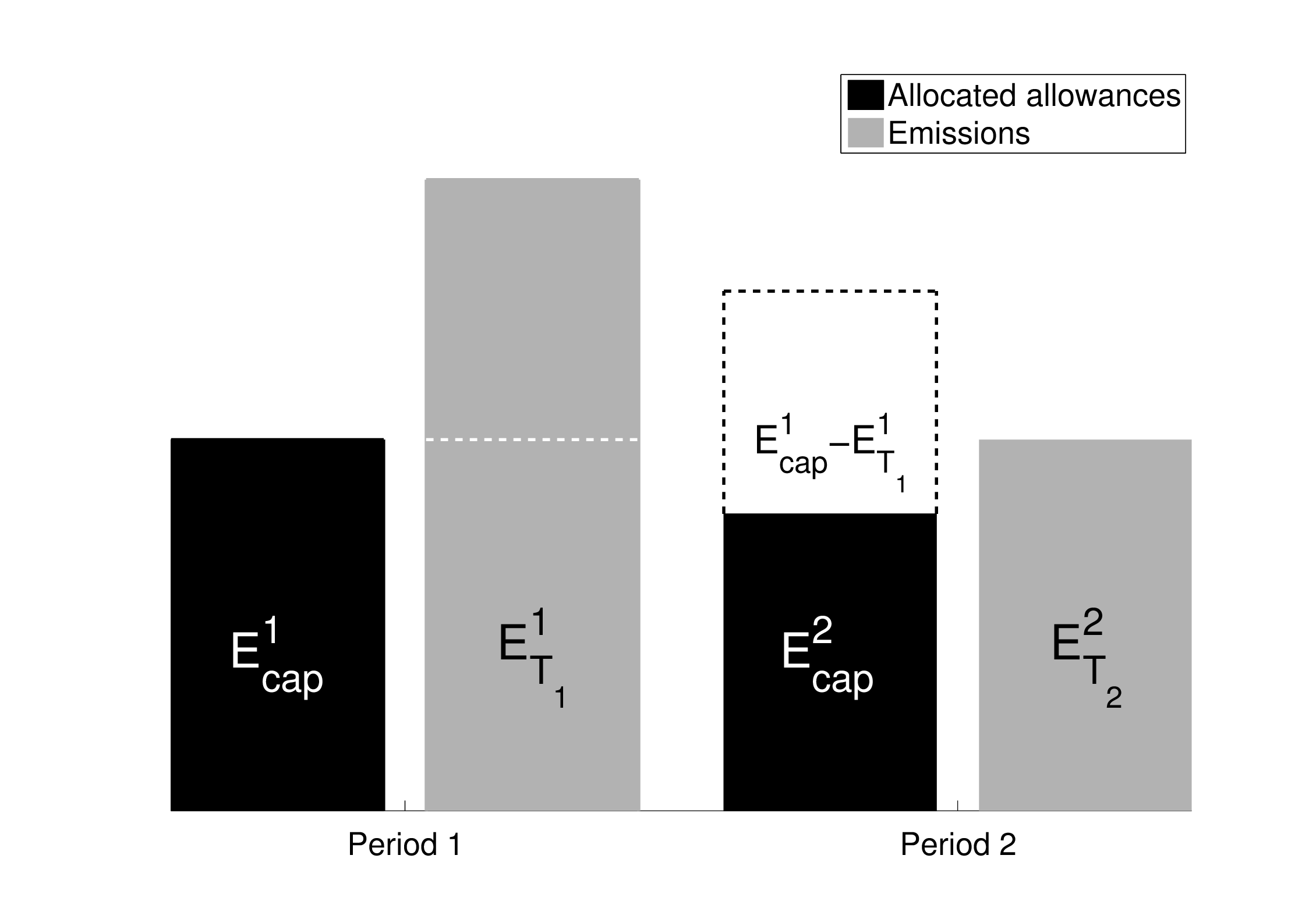}}
  \caption{Compliance period connecting mechanisms in an emissions market with two periods.}
 \label{fig:banking_withdrawal}
\end{figure}

These features imply that during the second period the aggregate supply of certificates now stems from two sources. First, the regulator issues a
number of permits $E_{\text{cap}}^2$ at the beginning of the period. Second, as explained above, a number of certificates are banked or withdrawn. The
aggregate supply of certificates during the second period is then given by
\begin{equation}\label{eq:aggregate_all_supply}
 \hat{E}_{\text{cap}}^2 = \left(E_{\text{cap}}^2 + E_{\text{cap}}^1 - E^1 \right)^+.
\end{equation}
Figure \ref{fig:banking_withdrawal} illustrates the banking and withdrawal mechanisms in the two-period market under consideration. In Figure
\ref{fig:banking_schematic} compliance at $t=T_1$ leads to the banking of a number $(E_{\text{cap}}^1-E^1)$ of certificates. The market is in
compliance at $t=T_2$ because of this additional supply of certificates. In Figure \ref{fig:withdrawal_schematic} noncompliance at $t=T_1$ leads to
the withdrawal of a number $(E^1-E_{\text{cap}}^1)$ of certificates. This leads to noncompliance at $t=T_2$ because of the decreased supply of
certificates during the second compliance period, even though cumulative emissions during the period $[T_1,T]$ are below the second-period cap.

The terminal condition $\phi_1$ for the allowance price in an emissions market with two compliance periods connected by the mechanisms of banking and
withdrawal now follows. Banking implies that in the event of compliance, that is, if $E_{T_1}<E^1_{\text{cap}}$, the value of the first period
allowance at time $T_1$ equals the value of the second period allowance at time $T_1$. In the event of noncompliance at time $T_1$ with
$E^1_{\text{cap}}\leq E_{T_1} < E^1_{\text{cap}} + E_{\text{cap}}^2$, the penalization of excess emissions and the withdrawal of certificates lead to
the first-period allowance certificate taking the value of the sum of the second-period certificate and the penalty. In the event of noncompliance at
time $T_1$ with $E_{T_1}\geq E^1_{\text{cap}} + E_{\text{cap}}^2$, the double penalization rule implies that the value of the first-period allowance
certificate equals the sum of $\Pi^1$ and $\bar{\Pi}^1$. Therefore, $\phi_1$  is given by
\begin{equation}\label{eq:allowance_terminal_cnd_multiple_cp_BW}
\phi_1(E_{T_1}) :=
\begin{cases}
 A^2_{T_1} &\text{for } 0 \leq E_{T_1}<E^1_{\text{cap}},\\
\Pi^1+A^2_{T_1} &\text{for } E^1_{\text{cap}}\leq E_{T_1} < E^1_{\text{cap}}+E_{\text{cap}}^2,\\
\Pi^1+\bar{\Pi}^1 &\text{for } E^1_{\text{cap}}+E_{\text{cap}}^2 \leq E_{T_1}\leq E_{\max}.
\end{cases}
\end{equation}

At time $T_2$, the terminal condition $\phi_2$ for the allowance price is now the same as in the one-period case with the exception that the aggregate
supply of certificates $\hat{E}^2_{\text{cap}}$ is used:
\begin{equation}\label{eq:allowance_terminal_cnd_second_cp_BW}
\phi_2\left(E_{T_2}\right) :=
 \begin{cases}
 0 	&\text{for } 0 \leq E_{T_2} < \hat{E}^2_{\text{cap}},\\
\Pi^2 	&\text{for } \hat{E}^2_{\text{cap}}\leq E_{T_2}\leq E_{\max}.
\end{cases}
\end{equation}
Note that the terminal condition $\phi_2$ uses the aggregate supply of certificates $\hat{E}_{\text{cap}}^2$, as defined in
\eqref{eq:aggregate_all_supply}, and is hence path-dependent. In particular, it depends on $E^1$, as mentioned earlier. In the context of pricing
futures contracts on allowance certificates in a two-period market, a similar terminal condition was introduced in \cite{rCarmona2010d}.

\subsubsection{Borrowing, banking, and withdrawal}
In addition to banking and withdrawal, consecutive compliance periods may also be connected by the borrowing mechanism. The effect of banking and
withdrawal at time $T_1$ is to increase the value of the first-period allowance certificate from zero to $A^2_{T_1}$ in the event of compliance (due
to the banking mechanism) and from $\Pi^1$ to $(\Pi^1+A^2_{T_1})$ or $(\Pi^1 + \bar{\Pi}^1)$ in the event of noncompliance (due to the withdrawal
mechanism). In contrast, the borrowing mechanism decreases the probability with which noncompliance occurs.

In an emissions market in which borrowing is allowed, firms may bring forward certificates from the second allocation $E_{\text{cap}}^2$ and use them
for compliance at time $T_1$. This does not affect the aggregate supply of certificates during the first compliance period, but whereas previously
noncompliance occurred when $E_{T_1}\geq E_{\text{cap}}^1$, this is no longer the case, as certificates from the second period may be borrowed to
supplement the aggregate supply during the first period. Therefore, noncompliance now occurs only if $E_{T_1}\geq E_{\text{cap}}^1 +
E_{\text{cap}}^2$, in which case the entire allocation of the second period must be borrowed and additional units of excess emissions are penalized at
the combined rate of $\Pi^1$ and $\bar{\Pi}^1$.

\begin{figure}[htbp]
  \centering
  \includegraphics[width=0.6\textwidth]{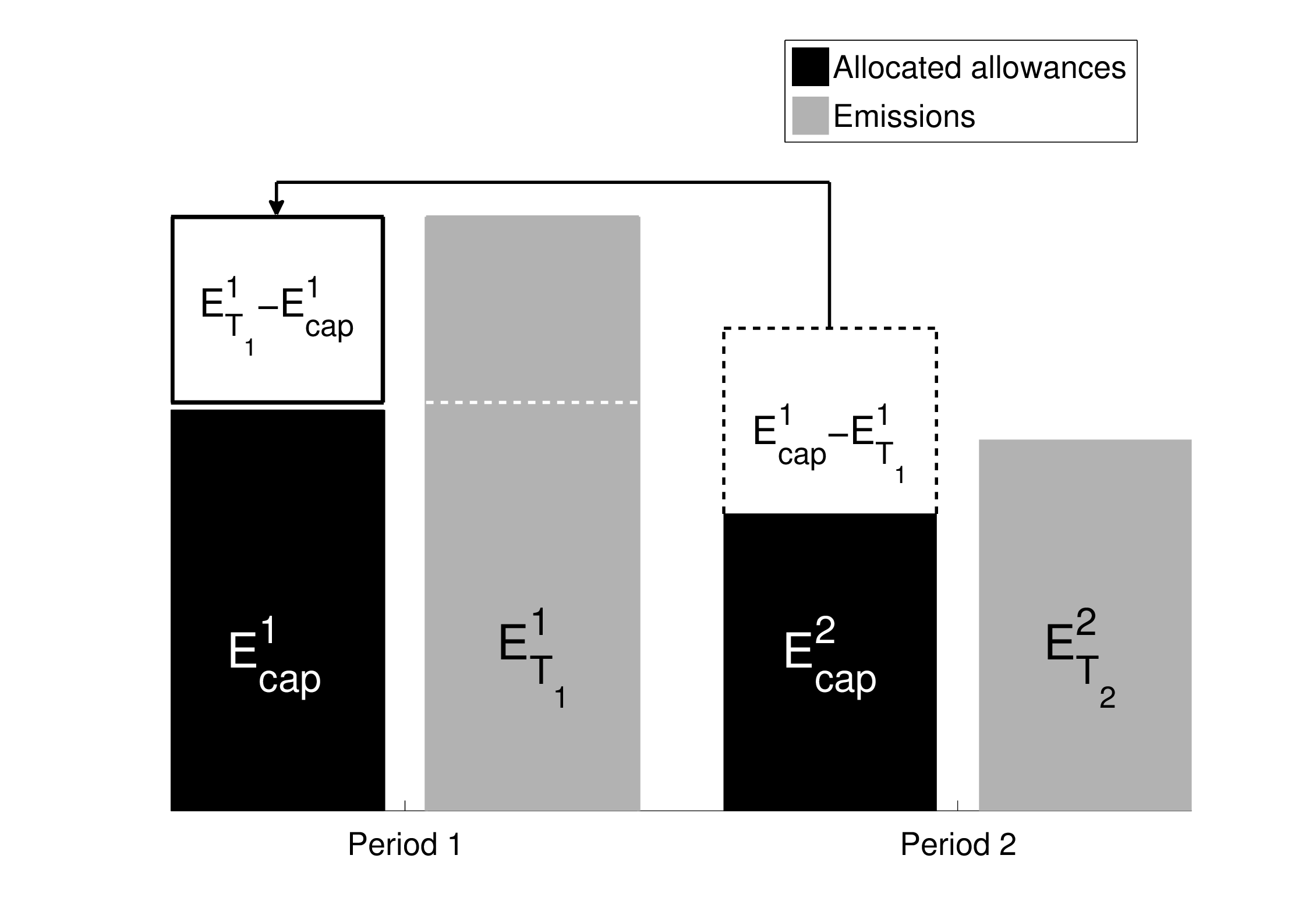}
  \caption{Borrowing mechanism in an emissions market with two periods.}
 \label{fig:borrowing}
\end{figure}

The borrowing mechanism is illustrated in Figure \ref{fig:borrowing}. Compliance at time $T_1$ is possible only by borrowing a number
$(E^1-E_{\text{cap}}^1)$ of certificates from the second compliance period. As a result these certificates are deducted from $E_{\text{cap}}^2$. In
our example this leads to noncompliance at time $T_2$.

Whereas $\phi_2$ continues to be given by \eqref{eq:allowance_terminal_cnd_second_cp_BW}, the terminal condition $\phi_1$ in an emissions market,
which connects subsequent compliance periods with the banking, borrowing, and withdrawal mechanisms, now follows and is given by
\begin{equation}\label{eq:allowance_terminal_cnd_multiple_cp_BBW}
\phi_1(E_{T_1}) :=
\begin{cases}
A^2_{T_1} &\text{for } 0 \leq E_{T_1}<E^1_{\text{cap}}+E_{\text{cap}}^2,\\
\Pi^1+\bar{\Pi}^1 &\text{for } E^1_{\text{cap}}+E_{\text{cap}}^2 \leq E_{T_1}\leq E_{\max}.
\end{cases}
\end{equation}

\section{Risk-neutral pricing of European derivatives}\label{str:option_pricing}
We now turn to the arbitrage-free pricing of European derivatives written on the allowance certificate within our model. For this purpose we work in
the emissions market of section \ref{str:allowance_pricing_one_cp} with one compliance period.

Our example of choice is a European call $(C_t(\tau))_{t\in[0,\tau]}$ with maturity $\tau$, where $0\leq \tau \leq T$, and strike $K\geq 0$, so that
its payoff is
\begin{equation*}\label{eq:call_payoff_one_period}
 C_\tau(\tau):=\left(A_\tau-K\right)^+.
\end{equation*}
We know from Assumption \ref{as:traded_assets} that, for $0\leq t \leq \tau \leq T$, the discounted call price
$\left(e^{-rt}C_t\right)_{t\in[0,\tau]}$ is a martingale under the measure $\mathbb{Q}$. Therefore, it is given as the discounted conditional
expectation of its terminal condition under this measure; i.e.,
\begin{equation*}
 C_t=e^{-r(\tau-t)}\tilde{\E}\left[\left.\left( A_\tau-K \right)^+\right|\mathcal{F}_t\right] \quad \text{for } 0\leq t \leq \tau.
\end{equation*}
As we argued previously for the allowance certificate, the discounted call price can be represented as an It\^{o} integral with respect to the
Brownian motion $(\tilde{W}_t)_{t\in[0,\tau]}$. It follows that
\begin{equation*}\label{eq:FBSDE_derivation_call}
 \Id \left(e^{-rt}C_t\right) = Z_t \Id \tilde{W}_t \quad \text{for } 0\leq t \leq \tau
\end{equation*}
and some $\mathcal{F}_t$-adapted process $(Z_t)_{t\in[0,\tau]}$.

Letting $C_t=v(t,D_t,E_t)$ for $0\leq t \leq \tau$, where
$v:[0,\tau]\times[0,\xi_{\max}]\times [0,E_{\max}]\mapsto\R_+$, we
find that $v$ satisfies
\begin{align}\label{eq:call_PDE}
 \mathcal{L} v &= 0 && \text{on } U,\ 0\leq t < \tau\nonumber\\
 v &= \left(\alpha(\tau,D,E)-K\right)^+ && \text{on } U,\ t=\tau,
\end{align}
where
\begin{equation*}
 \mathcal{L}:= \frac{\partial }{\partial t} + \frac{1}{2}\sigma_D^2(D) \frac{\partial^2 }{\partial D^2} +  \mu_D(D) \frac{\partial }{\partial D} +
\mu_E(\alpha(t,D,E),D) \frac{\partial }{\partial E} - r.
\end{equation*}
The relevant boundary conditions are discussed in section \ref{str:call_num}.

The key difference between the allowance certificate and the option pricing
problem is that the allowance price (representing the cost of carbon) has an
impact on the rate at which firms emit. This is reflected in the fact that
the drift of the cumulative emissions process depends on the price of the
allowance certificate but not on that of the option. Consequently, the FBSDE
\eqref{eq:allowance_FBSDE} is coupled, and the PDE \eqref{eq:allowance_PDE},
which describes the allowance price, is nonlinear, whereas
\eqref{eq:call_PDE} is linear.

\section{Asymptotics near expiry}
In this section (which is not in the original 
version of the paper~\cite{sHowison2012}) we
examine the asymptotic behaviour of the allowance price near expiry, where
the effect of the nonlinearity is most pronounced.

We recall that the allowance price $\alpha(t,D,E)$ satisfies the PDE
\[
\frac{\partial \alpha}{\partial t} 
+ \frac{1}{2}\sigma_D^2(D) \frac{\partial^2 \alpha}{\partial D^2} 
+  \mu_D(D) \frac{\partial \alpha}{\partial D} +
\mu_E(\alpha,D) \frac{\partial \alpha}{\partial E} - r\alpha =0
\]
with the terminal condition
\[
  \alpha(T,D,E) = \Pi\,\mathbb{I}_{[E_{\text{cap}},\infty)}(E)
\]
We examine the behavior near the discontinuity and as $t\to T$ by scaling
\[
T-t = \epsilon \tau, \qquad E - E_\mathrm{cap} = \epsilon \eta
\]
where $0<\epsilon\ll1$ is an artificial small parameter which helps with the
book-keeping.  We then
expand 
\[
\alpha(t,D,E)\sim \alpha_0(\tau,D,\eta)+ \epsilon \alpha_1(\tau, D, \eta)
+\cdots.
\]
At leading order, we obtain
\begin{equation}
-\frac{\partial \alpha_0}{\partial \tau} +\mu_E(\alpha_0,D)\frac{\partial \alpha_0}{\partial \eta}=0
\label{NewEq1}
\end{equation}
with
\[
\alpha_0(0, D, \eta) = \begin{cases} 0 \qquad & \eta <0,\\
                                     \Pi      & \eta \geq 0.
                       \end{cases}
\]
This is a standard problem and the
solution has an expansion fan between the lines $\eta = -\mu_E^-(D) \tau$ and
$\eta = -\mu_E^+(D)\tau$ where $\mu_E^-(D)=\mu_E(0,D)$ and
$\mu_E^+(D)=\mu_E(\Pi,D)$.  
We obtain
\[
\alpha_0(\tau, D, \eta) = \begin{cases} 0 \hfill  \eta &< -\mu_E^-(D) \tau \\
                              f(\xi;D) \qquad  -\mu_E^-(D)\tau &\leq  \eta \leq
                              -\mu_E^+(D)\tau \\
                                  \Pi \hfill \eta &> - \mu_E^+(D)\tau,
\end{cases}
\]
where $\xi = \eta/\tau$ and the function $f(\xi;D)$ is defined implicitly by
$\mu_E(f(\xi,D);D) = -\xi$, in which $D$ appears only as a parameter.

\begin{figure}[htbp]
  \centering
  \input{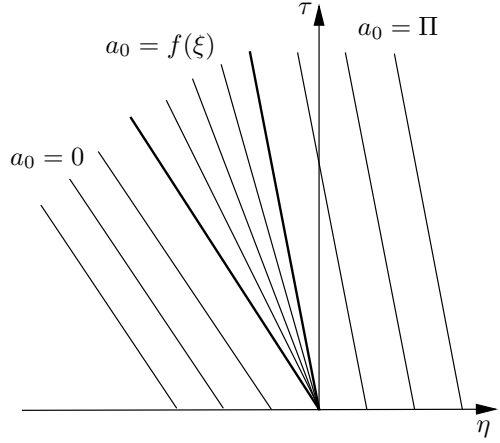}
  \caption{Sketch of characteristics for leading-order approximation 
           near expiry.}
 \label{fig:NewSec.1}
\end{figure}
  
\bigskip


The characteristics of~(\ref{NewEq1}) are sketched in
Figure~\ref{fig:NewSec.1} with $D$ fixed, so that this is a two-dimensional
slice through a three-dimensional expansion fan which varies parametrically
with $D$.  It is important to note that the fan borders (the bold lines) vary
monotonically in $D$, because $\mu_E(\cdot,D)$ is strictly increasing in
$D$ by assumption. This allows diffusion in the $D$-direction to
smooth the gradient discontinuities that occur across these lines.  Let us look at the
lower border---the expansion around the upper border is similar---and 
construct an inner-layer expansion by changing variables to
\[
\eta=-\mu_E^-(D)\tau + \epsilon^\frac12 y, \qquad 
\alpha_0(\tau,D,\eta)=\epsilon^\frac12
a_0(\tau,D,y).
\]
The term $\frac12\sigma_D^2(D)\partial^2\alpha/\partial D^2$
in the original PDE now enters at the same order 
 as the leading-order terms retained above. We therefore obtain, at leading
 order, 
the nonlinear parabolic PDE
\begin{equation}
\frac{\partial a_0}{\partial \tau}=
\frac12\sigma_D^2\left(\tau \frac{\mathrm{d} \mu_E^-}{\mathrm{d} D}\right)^2
\frac{\partial a_0}{\partial y} + \frac{\mathrm{d} \mu_E^-}{\mathrm{d} D}a_0\frac{\partial a_0}{\partial y}
\label{NewEq.2}
\end{equation}
with the initial condition
\[
a_0(0,D,y) = \Pi\max(y,0)
\]
which is the initial gradient discontinuity. 
(It is this condition that dictates the
scaling for $\alpha_0$ above.)

There is no similarity solution to~(\ref{NewEq.2}), 
but its short-time (small $\tau$) behavior is of the form 
\[
a_0 \sim \tau^\frac32 g(y/\tau^\frac32)\left(1+O(\tau)\right)
\]
where $g(z)$ (writing $z = y/\tau^\frac32$) satisfies
\[
\frac32\left(g - z\frac{\mathrm{d}g}{\mathrm{d}z}\right)
=\frac12\Sigma^2_D\frac{\mathrm{d}^2g}{\mathrm{d}z^2},
\]
in which $\Sigma_D:=\sigma_D\mathrm{d}\mu_E^-/\mathrm{d}D$. The initial
condition forces $g(z)\to 0$ as $z\to -\infty$ and $g(z)\sim \Pi z$ as
$z\to\infty$, and the solution is 
\[
g(z)= \Pi\left( \zeta N(\zeta) - \mathrm{e}^{-z^2/2}/\sqrt{2\pi}\right)
\]
in which $N(\cdot)$ is the cumulative density of the standard Normal
distribution and  
$\zeta = z\sqrt{3}/\Sigma_D$.

The far-field behavior of~(\ref{NewEq.2}) is more straightforward; a balance
of the first and last terms, namely  
\[
\frac{\partial a_0}{\partial \tau}\sim\frac{\mathrm{d} \mu_E^-}{\mathrm{d} D}a_0\frac{\partial a_0}{\partial y},
\]
leads to the approximate similarity solution 
\[
a_0(z,\tau)\sim -\frac{1}{\mathrm{d} \mu_E^-/\mathrm{d} D}\frac{z}{\tau} 
\]
and this is readily shown to match with the inner expansion of the
expansion-fan solution.

\subsection{The probability that $E_T=E_\mathrm{cap}$}
The incentives offered by the market in our model pull in two ways. On the
one hand, in the absence of a carbon penalty, Business-As-Usual offers the
cheapest (but the dirtiest) means of production---this is what the bid stack
delivers---and in general reducing
emissions increases prices. On the other hand, the threat of paying the
penalty should act to keep total emissions below $E_\mathrm{cap}$. It is
therefore interesting to ask whether this tension leads to a non-zero
probability that $E_t$ tends to $E_\mathrm{cap}$ from below as $t\to
T$. Carmona \& Delarue show (in a different but related model) that precisely
this occurs using techniques of Malliavin calculus (cf. \cite{rCarmona2011b}). We now
give a heuristic argument why this occurs (in the framework of our
model). The non-zero probability that $E_t$ reaches $E_{\text{cap}}$
from below can be interpreted informally as saying that the feedback
and the associated nonlinearity combine to achieve the largest
possible proportion of total emission trajectories that just miss
having to pay the penalty.

To make this more concrete, we start by noting that, as $\mathrm{d} E_t = \mu_E(A_t,D_t)\,\mathrm{d}t$,
continuity of the various functions and processes involved means that paths
of $E_t$ are $C^1$. Hence, we expect paths that lie in the expansion fan
in Figure~\ref{fig:NewSec.1} for $t$ close to $T$ to reach $E=E_\mathrm{cap}$
($\eta = 0$) tangentially to one of the characteristics in the fan (extended
in the $D$-direction). This suggests that a whole collection of paths are
forced to the single point $E_T=E_\mathrm{cap}$ 
and that the probability mentioned above is non-zero.

A more quantitative argument is as follows. First note that, for any $0\leq
\mathcal{E}\leq E_\mathrm{max}$, 
\[
P_t:=\mathbb{P}\left(E_T > \mathcal{E} | E_t, D_t\right) 
= \mathbb{E}\left(\mathbb{I}_{(\mathcal{E},\infty)}(E_T)|E_t,D_t\right).
\] 
This conditional expectation satisfies the \emph{linear} 
equation~(\ref{eq:call_PDE}) with $r$ set to zero, namely
\begin{equation}
\frac{\partial P}{\partial t} 
+ \frac{1}{2}\sigma_D^2(D) \frac{\partial^2 P}{\partial D^2} +  
\mu_D(D) \frac{\partial P}{\partial D} +
\mu_E(\alpha(t,D,E),D) \frac{\partial P}{\partial E} =0.
\label{eq:LinProb}
\end{equation}
This is well known; one way to see it in a financial context is to write
\[
P_t
= \mathrm{e}^{r(T-t)}\times\mathbb{E}
\left( \mathrm{e}^{-r(T-t)}\mathbb{I}_{(\mathcal{E},\infty)}(E_T)|E_t,D_t\right)
\]
and to note that the second multiplicand above is the value of a
derivative contract, which in the finance literature is known as a digital call option on
$E_T$ with strike $\mathcal{E}$, which therefore
satisfies~(\ref{eq:call_PDE}) with $r$ 
included; substitution gives the result.

If we  define
\[
P_t^\pm = \mathbb{P}\left(E_T > E_\mathrm{cap}\pm\delta | E_t, D_t\right),
\]
then 
\[
 \mathbb{P}\left(E_T = E_\mathrm{cap}| E_t, D_t\right)
=\lim_{\delta\to 0}\left(P_t^--P_t^+\right)
\]
Now consider the local expansion of~(\ref{eq:LinProb}) near
$E=E_\mathrm{cap}$, $t=T$ as above: at leading order
in an expansion $P(t,E,D)\sim P_0(\tau, \eta, D)+\cdots$, we see that
\[
 -\frac{\partial P_0}{\partial \tau} +\mu_E(\alpha_0(\tau,\eta,D),D)\frac{\partial P_0}{\partial \eta}=0.
\]
This linear equation has the same characteristics as its nonlinear cunterpart
for $\alpha_0$, and $P_0$ is constant along them. Thus,
\begin{itemize}
\item $P_0^+$ is equal to 1 on all characteristics starting from $E >
  E_\mathrm{cap} + \delta$, that is from just to the right of the right-hand bold
  line in Figure~\ref{fig:NewSec.1} and upwards, 
and zero on all the others (including
  the expansion fan).
\item $P_0^-$ is equal to zero on all characteristics starting from $E\leq
  E_\mathrm{cap} - \delta$, that is, from just to the left of the left-hand
  bold line and downwards, and equal to 1 on all the others (including the
  expansion fan).\footnote{An expansion similar to that above shows that
    there is a diffusive smoothing of the discontinuity in $P^\pm$ for
    $\tau>0$. Using the same notation as above, the solution is less
    complicated,  being described by the equation
\[
\frac{\partial p}{\partial \tau}=\frac12\Sigma_D^2\tau^2\frac{\partial p}{\partial z}
\]
with a typical solution taking the form $N(z\sqrt{3}/\Sigma_D)$, where
again $z=y/\tau^\frac32$. }
\end{itemize} 
It is now apparent that $P_t^--P_t^+$ tends to 1 in the expansion fan; and
consequently this non-zero value is propagated out into the whole domain by
the equation~(\ref{eq:LinProb}). That is, we have a non-zero probability that
total emissions just reach $E_\mathrm{cap}$ from below (as indicated by the
orientation of the expansion fan).

\section{Numerical analysis}\label{str:num_analysis}
This section is dedicated to the numerical analysis of the model. We illustrate the dependency of allowance prices on demand and the cumulative
emissions and compare prices in the setting of a single-period market to those implied by multiperiod markets. Further, we demonstrate the dependence
structure of a European option written on the allowance certificate.

\subsection{Concretizing the model}
We begin by specifying the functions and parameters in the model.

\subsubsection{Functional form of the bid and the emissions stack}
We take the business-as-usual bid stack to be of the form
\begin{equation*}
b^{\text{BAU}}(\xi):= \underline{b} + \left(\frac{\overline{b}-\underline{b}}{\xi_{\max}^{\theta_1}}\right)\xi^{\theta_1} \quad \text{for } 0\leq \xi
\leq \xi_{\max},
\end{equation*}
where $\underline{b}, \overline{b} \geq 0$ and $2 < \theta_1 < \infty$. With this choice $b^{\text{BAU}}$ is strictly convex and strictly increasing
on its domain of definition. The parameters $\underline{b}$ and $\overline{b}$ correspond to the minimum and maximum prices of electricity the model
can produce. Because the range of allowed bids and typically observed market prices in many auction-based electricity markets is well known, these are
relatively easy to infer in practice. The parameter $\theta_1$ controls the steepness of the stack and in particular how quickly marginal costs of
generators increase.

Similarly, we take the marginal emissions stack to be of the form
\begin{equation*}
e(\xi):= \overline{e} - \left(\frac{\overline{e}-\underline{e}}{\xi_{\max}^{\theta_2}}\right)\xi^{\theta_2} \quad \text{for } 0\leq \xi \leq
\xi_{\max},
\end{equation*}
where $\underline{e}, \overline{e} \geq 0$ and $0 \leq \theta_2 < 1$. With this definition also $e$ is strictly convex and decreasing on its domain of
definition. The parameters $\underline{e}$ and $\overline{e}$ correspond to the minimum and maximum marginal emissions rates in the market. In a market
with coal and gas generators only and under the reasonable assumption that coal is the more emission-intensive technology than gas, $\overline{e}$
would represent the marginal emissions rate of coal and $\underline{e}$ that of gas. The parameter $\theta_2$ controls the fuel mix in the market. The
smaller the value of $\theta_2$, the smaller the proportion of the market capacity that is served by the pollution-intensive technology.

Clearly $b^{\text{BAU}}$ and $e$ satisfy the assumptions in Definitions \ref{def:BAU_stack} and \ref{def:ems_stack}. Further, since a linear
combination of strictly convex functions is also strictly convex, so is the function $g$. Therefore, Assumption \ref{as:b_measure_zero} is also
satisfied. Moreover, we note that for this choice of bid and emissions stack the set $S_p(\cdot,\cdot)$ is always of the form $[\xi_1,\xi_2]$ for
$0\leq \xi_1 \leq \xi_2 \leq \xi_{\max}$.

\subsubsection{The demand process}\label{str:demand_jacobi}
We specify that under $\mathbb{\tilde{P}}$ the process $(D_t)$ follows the stochastic differential equation
\begin{equation}\label{eq:demand_coefficients}
 \Id D_t = -\eta\left(D_t-\bar{D}\right) \Id t + \sqrt{2\eta \bar{\sigma}_D D_t\left(\xi_{\max} - D_t\right)} \Id \tilde{W}_t, \qquad D_0=d \in
(0,\xi_{\max}),
\end{equation}
where $\overline{D}, \eta, \bar{\sigma}_D > 0$. With this definition $(D_t)$ is a Jacobi diffusion process; it has a linear, mean-reverting, drift
component and degenerates on the boundary. Moreover, subject to $\overline{D}\in(0,\xi_{\max})$ and $\min(\overline{D},\xi_{\max}-\overline{D})\geq
\xi_{\max}\bar{\sigma}_D$, the process remains within the interval $(0,\xi_{\max})$; its stationary distribution is a beta distribution, and its mean
is given by $\bar{D}$ (cf.\ \cite{jForman2008}).

\subsubsection{Choice of parameters}
Tables \ref{tab:ex_bid_ems_stacks}, \ref{tab:ex_demand}, and \ref{tab:ex_market} summarize the parameter values used for the numerical study that follows.
We note that they do not correspond to a particular example of an electricity market, but they can be considered representative of a medium-sized
market whose fuel mix predominantly consists of coal and gas generators.

Table \ref{tab:ex_bid_ems_stacks} specifies the parameters for the bid and the emissions stack. Using \eqref{eq:cumulative_emissions_process} now with
$A_t=0$ and $D_t = \xi_{\max}$ for $0\leq t \leq T$, and with the assumption that there are $24\times 365$ production hours in the year, we find that
$E_{\max}=1.6519\times 10^8$.

\begin{table}[h]
\footnotesize
 \caption{Parameters for the bid and emissions stack.}\label{tab:ex_bid_ems_stacks}
 \begin{center}
  \begin{tabular}{cccccccc}
    \toprule
     $\overline{b}$ & $\underline{b}$ & $\theta_1$ & $\overline{e}$ & $\underline{e}$ & $\theta_2$ & $\kappa$ & $\xi_{\max}$\\
    \midrule
    $200$ & $0$ & $10$ & $1.2$ & $0.4$ & $0.4$ & $8760$ & $30000$\\
    \bottomrule	
  \end{tabular}
 \end{center}
\end{table}

The parameters relating to demand are given in Table \ref{tab:ex_demand}.

\begin{table}[h]
\footnotesize
 \caption{Parameters for the demand process and the risk-free rate.}\label{tab:ex_demand}
 \begin{center}
  \begin{tabular}{cccc}
    \toprule
    $\eta$ & $\bar{D}$ & $\bar{\sigma}_D$ & $r$\\
    \midrule
    $10$ & $21000$ & $0.05$ & $0.05$\\
    \bottomrule	
  \end{tabular}
 \end{center}
\end{table}

Calculating the cumulative emissions now for $A_t=0$ and demand at its mean level $D_t = \bar{D}$ for $0\leq t \leq T$,  we find that $E_T =
1.2961\times 10^8$. This leads us to choose the cap slightly below this level in order to incentivize a reduction in emissions. The parameters
characterizing the emissions trading scheme are given in Table \ref{tab:ex_market}. We note that here time is measured in years.

\begin{table}[h]
\footnotesize
 \caption{Parameters characterizing the emissions trading scheme.}\label{tab:ex_market}
 \begin{center}
  \begin{tabular}{ccc}
    \toprule
    $E_{\text{cap}}$ & $\Pi$ & $T$\\
    \midrule
    $1.17\times 10^8$ & $100$ & $1$\\
    \bottomrule	
  \end{tabular}
 \end{center}
\end{table}

\subsection{The allowance price value function}\label{str:al_num}
We present the necessary boundary conditions for the allowance price valuation equation and discuss its solution. First, this is done in the setting
of an emissions market with one compliance period, and second we consider two periods connected by either banking and withdrawal or banking, borrowing, and
withdrawal.

With regards to problem \eqref{eq:allowance_PDE} (in the case of one period) or \eqref{eq:allowance_PDE_multiple_cp} (in the case of two periods),
the following questions arise: At which points of the boundary do we need to specify boundary conditions in addition to the terminal condition, and what conditions
make sense given the original stochastic problem \eqref{eq:allowance_FBSDE} or \eqref{eq:allowance_FBSDE_multi_period}? The former question is
answered by considering the Fichera function $f$ (cf.~\cite{oOleinik1973}). Defining $n:=(n_D,n_E)$ to be the inward normal vector to the boundary,
Fichera's function for the operator $\mathcal{N}$ (and $\mathcal{L}$) reads
 \begin{equation*}
 f(t,D,E):= \left(\mu_D(D)-\frac{1}{2}\frac{\partial }{\partial D}\sigma_D^2(D)\right)n_D + \mu_E(\alpha(t,D,E),D)n_E \quad \text{on } \partial U_T
 \end{equation*}
 (or $\partial U_{T_i}$). In the case when the coefficients $\mu_D$ and $\sigma_D$ are of the form prescribed in \eqref{eq:demand_coefficients} we
find that
 \begin{equation*}
 f(t,D,E) =\eta\left(\left(\bar{D}-\bar{\sigma}_D \xi_{\max}\right) + \left(2\bar{\sigma}_D-1\right)D\right)n_D + \mu_E(\alpha(t,D,E),D)n_E \quad
\text{on } \partial U_T
 \end{equation*}
 (or $\partial U_{T_i}$). At points of the boundary, where $f\geq 0$, information is outward flowing and no boundary conditions have to be specified;
at those points where $f < 0$ the information is inward flowing and boundary conditions are necessary. Considering the parts of the boundary
corresponding to $D=0$ and $D=\xi_{\max}$, we find that $f\geq 0$ if and only if $\min(\overline{D},\xi_{\max}-\overline{D})\geq
\xi_{\max}\bar{\sigma}_D$, which is the same condition prescribed in section \ref{str:demand_jacobi} to guarantee that the Jacobi diffusion stays
within the interval $(0,\xi_{\max})$. At points of the boundary corresponding to $E=0$, we find that $f\geq 0$ always. On the part of the boundary on
which $E=E_{\max}$, $f< 0$ except at the point $(D,E)=(0,E_{\max})$, where $f=0$, an ambiguity which could be resolved by smoothing the domain.
Therefore, no boundary conditions are necessary except when $E=E_{\max}$. The nature of the condition at this part of the boundary depends on whether
we consider a market with one compliance period or multiple periods, and we specify it in the relevant sections below.

Given values for the demand for electricity and the cumulative emissions, the valuation equation representing the allowance pricing problem determines
the arbitrage-free price of an allowance certificate. We illustrate this dependency by solving the PDE numerically, using
the finite difference scheme explained in Appendix \ref{ap:num_scheme_allowance}.

\subsubsection{One compliance period}
The boundary condition at $E=E_{\max}$ takes the form
\begin{equation}
   \alpha(t,D,E) = e^{-r(T-t)}\Pi, \quad [0,T)\times (0,\xi_{\max})\times \{E=E_{\max}\}.\label{eq:allowance_PDE_bd2}
\end{equation}

The condition \eqref{eq:allowance_PDE_bd2} follows from the fact that, as soon as the cumulative emissions surpass the cap, every additional tonne of
CO$_2$ is penalized at a rate $\Pi$ at time $T$.

\begin{figure}[htbp]
  \centering
  \subfloat[$t=T/2$.]{\label{fig:allvalsurf_one_cp_t1}\includegraphics[width=0.5\textwidth]{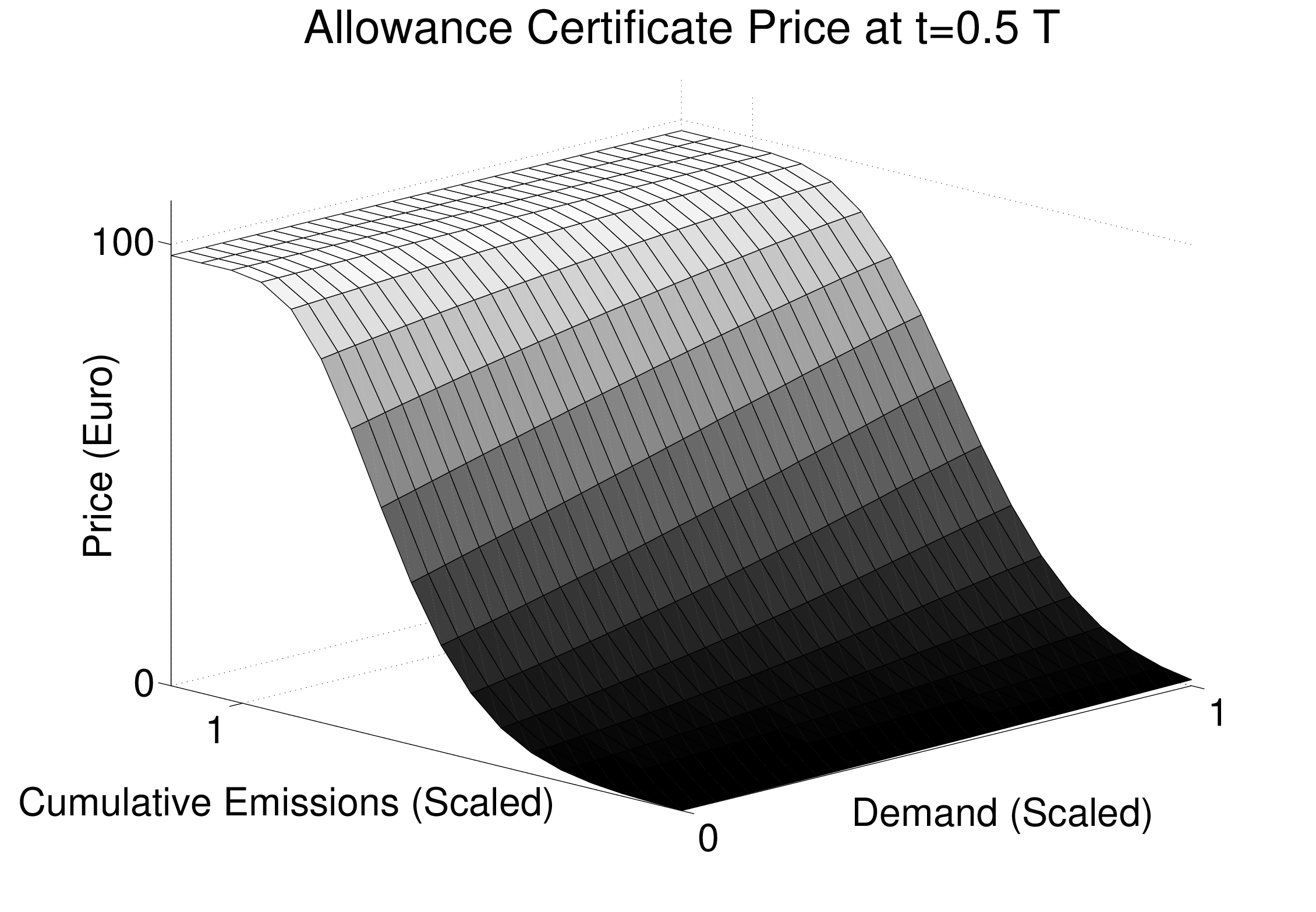}}
  \subfloat[$t=T$.]{\label{fig:allvalsurf_one_cp_tT}\includegraphics[width=0.5\textwidth]{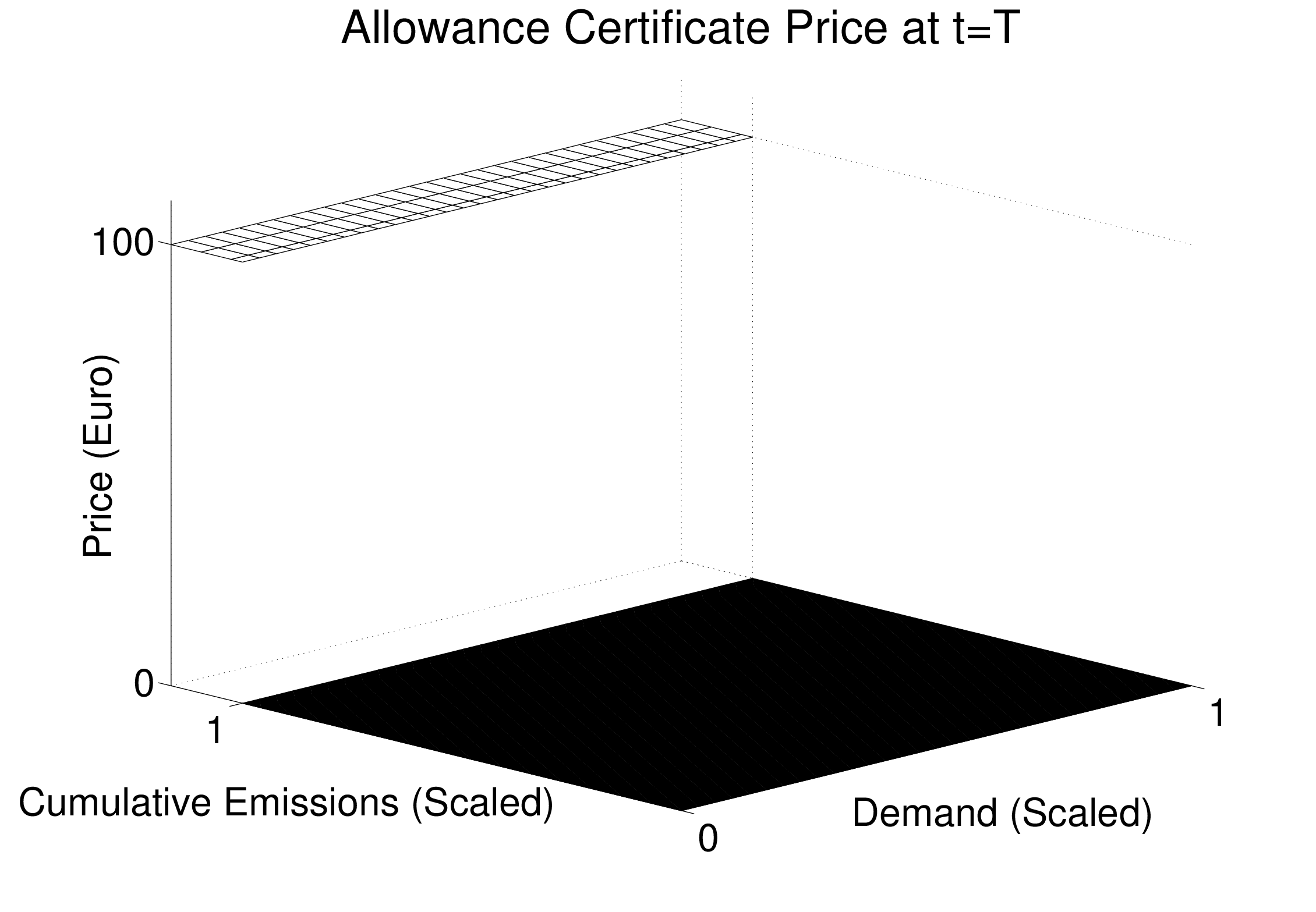}}
  \caption{The two plots show the price of an allowance certificate, in an emission market with one compliance period, at different times up to expiry
($\Pi=100$\euro).}
 \label{fig:allvalsurf_one_cp}
\end{figure}

The numerical results are displayed in Figure \ref{fig:allvalsurf_one_cp}. At time $t=T/2$, the allowance price depends on the cumulative emissions to
date and the current level of demand, as shown in Figure \ref{fig:allvalsurf_one_cp_t1}. For each fixed level of emissions $E=E_{T/2}$,
$\alpha(T/2,D,E_{T/2})$ is increasing in $D$. This makes intuitive sense, since for higher levels of demand, the corresponding market emissions rate
is greater, and consequently it is more likely that the cap will be reached. Similarly, fixing $D=D_{T/2}$ results in $\alpha(T/2,D_{T/2},E)$ being an
increasing function of $E$. In particular, we can think of the current level of cumulative emissions determining an interval for the allowance price
and the demand for electricity setting the exact price within this interval. Further, we notice that the allowance price equals the discounted
penalty if cumulative emissions exceed the cap. At the end of the compliance period, $\alpha$ is given by the terminal condition
\eqref{eq:allowance_terminal_cnd}. Figure \ref{fig:allvalsurf_one_cp_tT} reflects the digital nature of the price at this time and its independence of
$D$.

\subsubsection{Multiple compliance periods: Banking and withdrawal}

We illustrate the framework introduced in this section for the case of two consecutive compliance periods. We determine the prices of the first- and
second-period allowance certificates as functions of demand and cumulative emissions. For $i=2$, we solve the PDE
\eqref{eq:allowance_PDE_multiple_cp} with the terminal condition $\phi_2$ given by \eqref{eq:allowance_terminal_cnd_second_cp_BW}.
Further, the boundary condition at $E=E_{\max}$ takes the form
\begin{equation}\label{eq:eq:allowance_PDE_boundary_cond_second_cp_BW}
 \alpha_2\left(t,D,E\right) = e^{-r(T_2-t)}\Pi^2, \quad  [T_1,T_2)\times (0,\xi_{\max})\times \{E=E_{\max}\}.
\end{equation}

The problem is equivalent to the one-period pricing problem with the exception that the aggregate supply of certificates $\hat{E}_{\text{cap}}^2$
depends on $E^1$, the level of the cumulative emissions at the end of the first compliance period. As a result, the price of the second-period
allowance certificate depends not only on the current values of $t,\ D$, and $E$ but also on $E^1$, i.e., $\alpha_2=\alpha_2(\cdot,\cdot,\cdot ;E^1)$.
We then solve \eqref{eq:allowance_PDE_multiple_cp} for $i=1$ with the terminal condition $\phi_1$ of the form
\eqref{eq:allowance_terminal_cnd_multiple_cp_BW}, where $A^2_{T_1}=\alpha_2(T_1,D,0;E)$. The boundary condition at $E=E_{\max}$ now reads
\begin{equation}\label{eq:allowance_PDE_boundary_cond_first_cp_BW}
 \alpha_1\left(t,D,E\right) = e^{-r(T_1-t)}\left(\Pi^1+\bar{\Pi}^1\right), \quad  [0,T_1)\times (0,\xi_{\max})\times\{E=E_{\max}\}.
\end{equation}

\begin{figure}[htbp]
  \centering
  \subfloat[$t=T_1/2$.]{\label{fig:allvalsurf_two_cp_BW_t1}\includegraphics[width=0.5\textwidth]{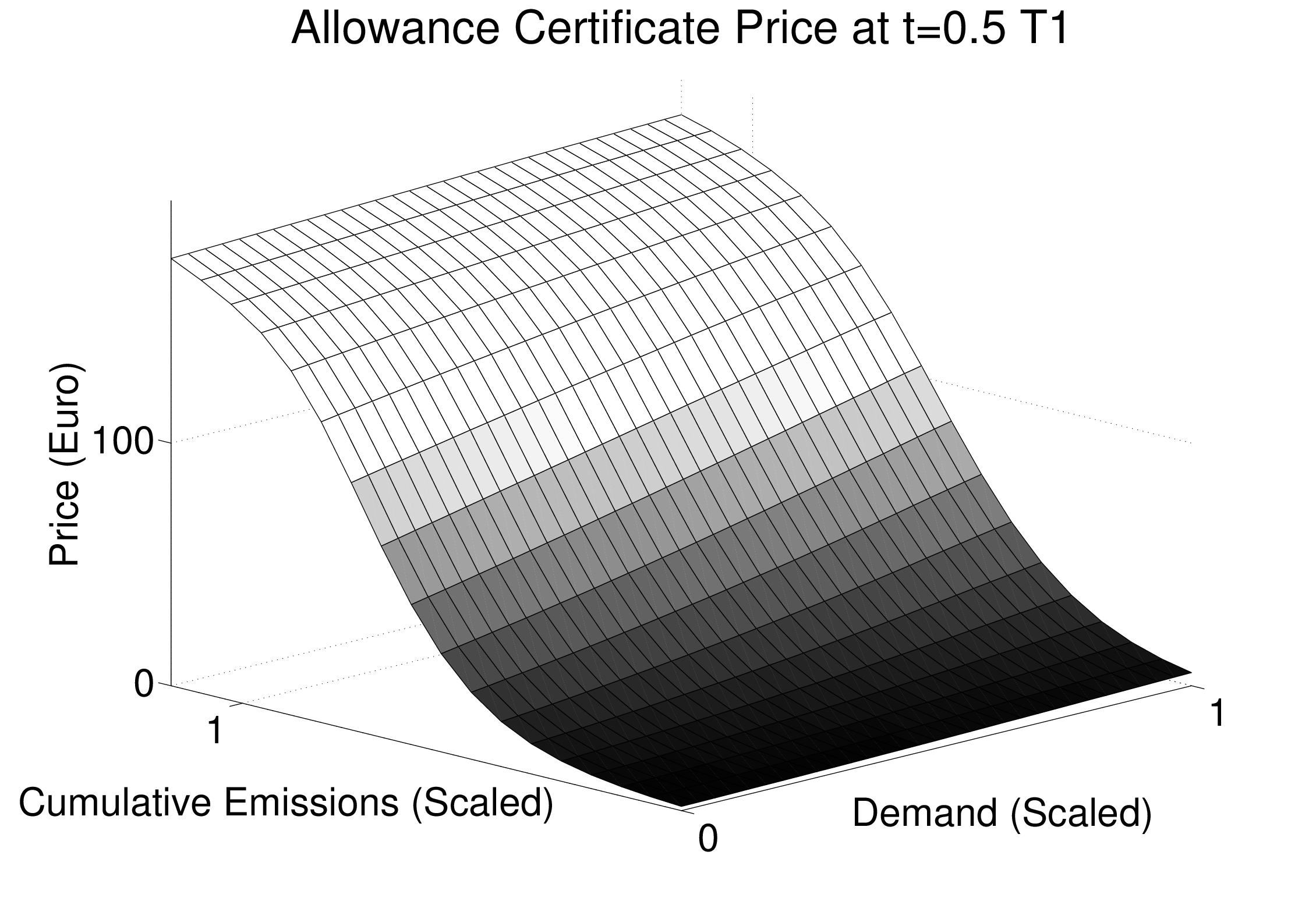}}
  \subfloat[$t=T_1$.]{\label{fig:allvalsurf_two_cp_BW_tT1}\includegraphics[width=0.5\textwidth]{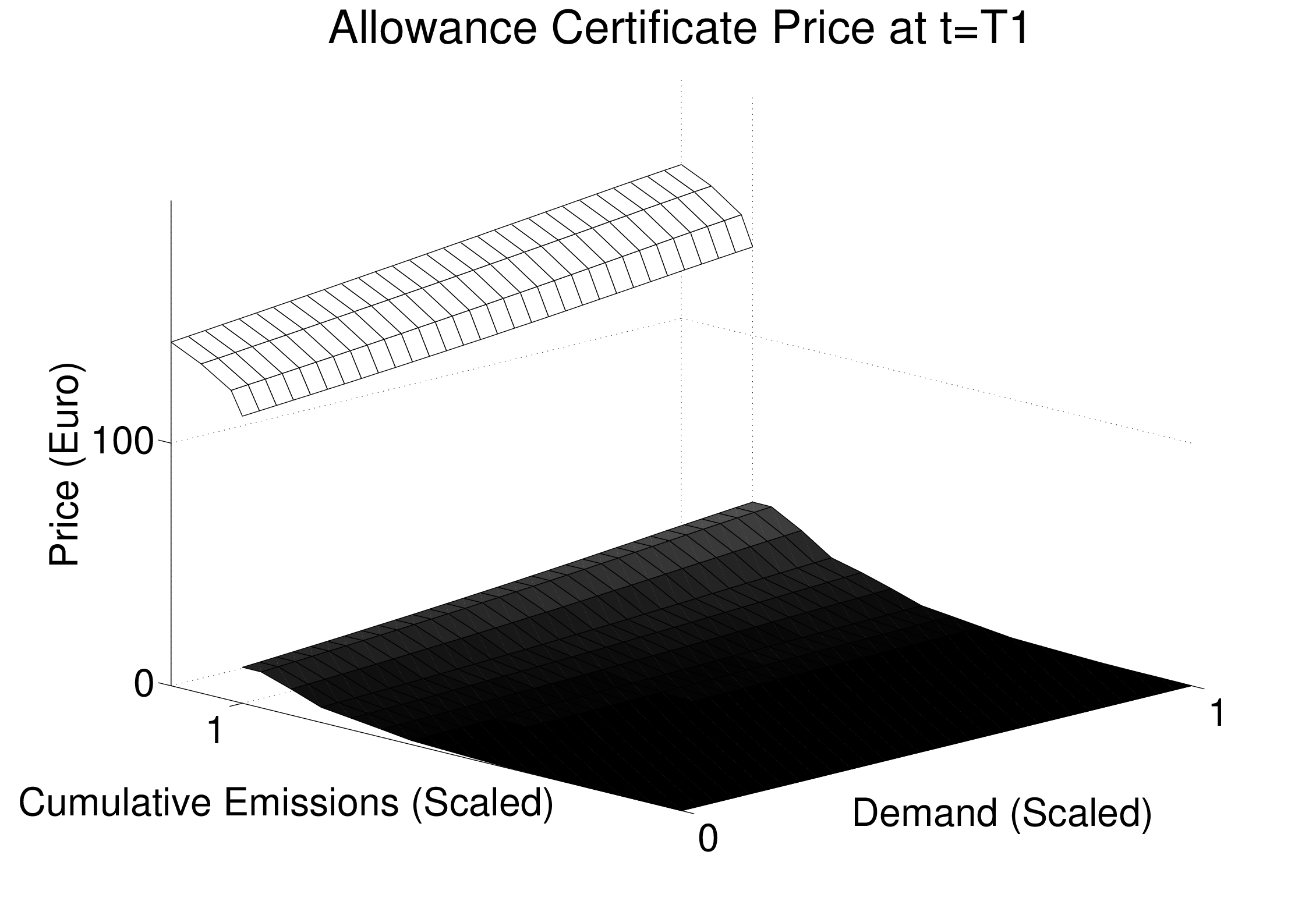}}
  \caption{The two plots show the value of the first-period allowance certificate $(A^1_t)_{t\in[0,T_1]}$ at different times up to compliance in an
emissions market with two compliance periods, which are connected by the banking and withdrawal mechanisms ($\Pi^1=100$\euro).}
\label{fig:allvalsurface_two_cp_BW}
\end{figure}

Figure \ref{fig:allvalsurf_two_cp_BW_t1} plots the value of the allowance certificate at time $t=T_1/2$. The effects of banking and withdrawal become
very clear as the value of the certificate exceeds the penalty for a sufficiently high level of cumulative emissions. At the end of the compliance
period $\alpha$ is given by the terminal condition \eqref{eq:allowance_terminal_cnd_multiple_cp_BW}, as shown in Figure
\ref{fig:allvalsurf_two_cp_BW_tT1}. Concerning the price behavior of the second-period certificate, we note that it mirrors the one-period model with
the initial allocation $E_{\text{cap}}$ replaced by $\hat{E}_{\text{cap}}^2$.

\subsubsection{Multiple compliance periods: Borrowing, banking, and withdrawal}
\begin{figure}[ht]
  \centering
  \subfloat[$t=T_{1/2}$.]{\label{fig:allvalsurf_two_cp_BBW_t1}\includegraphics[width=0.5\textwidth]{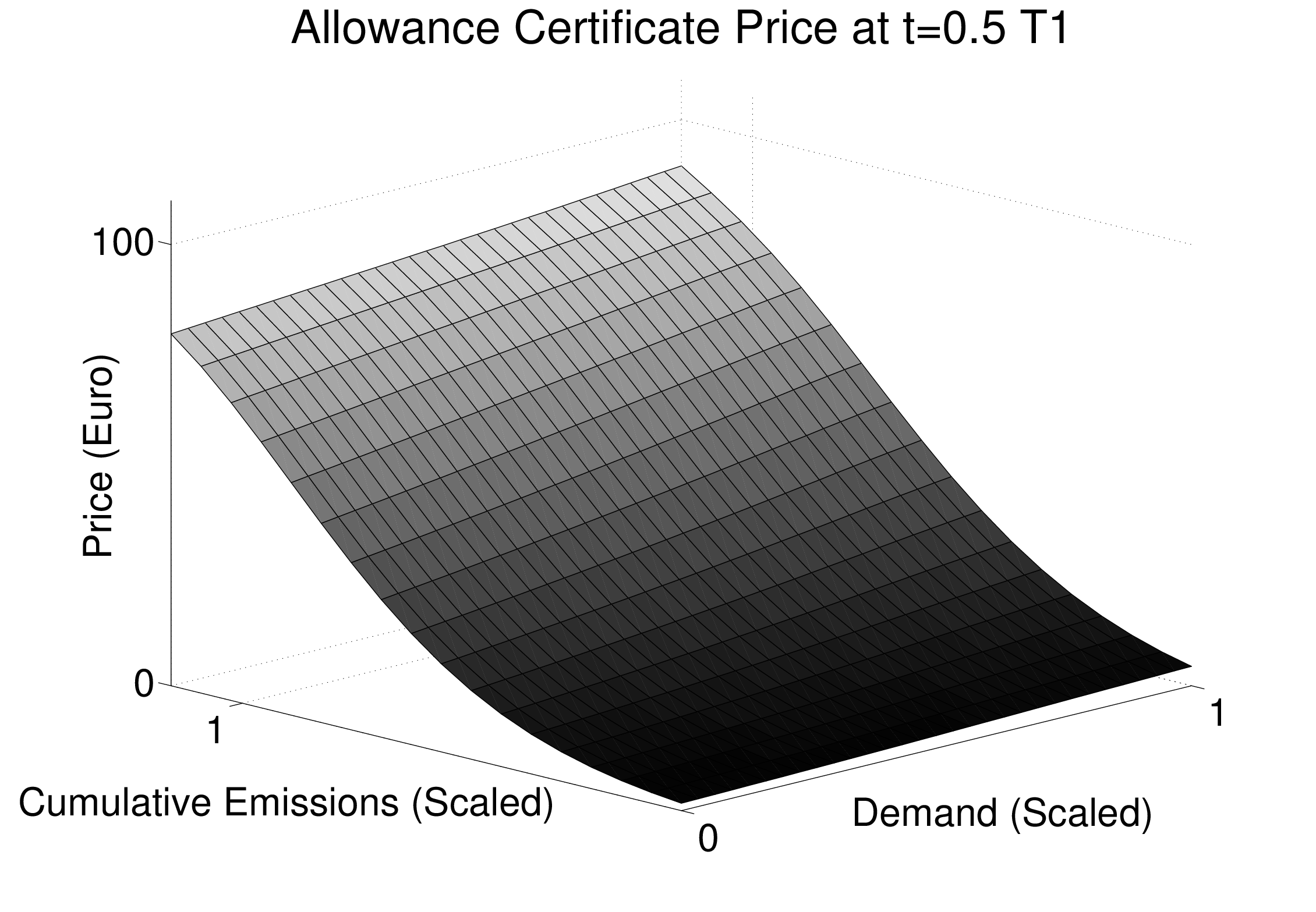}}
  \subfloat[$t=T_1$.]{\label{fig:allvalsurf_two_cp_BBW_tT1}\includegraphics[width=0.5\textwidth]{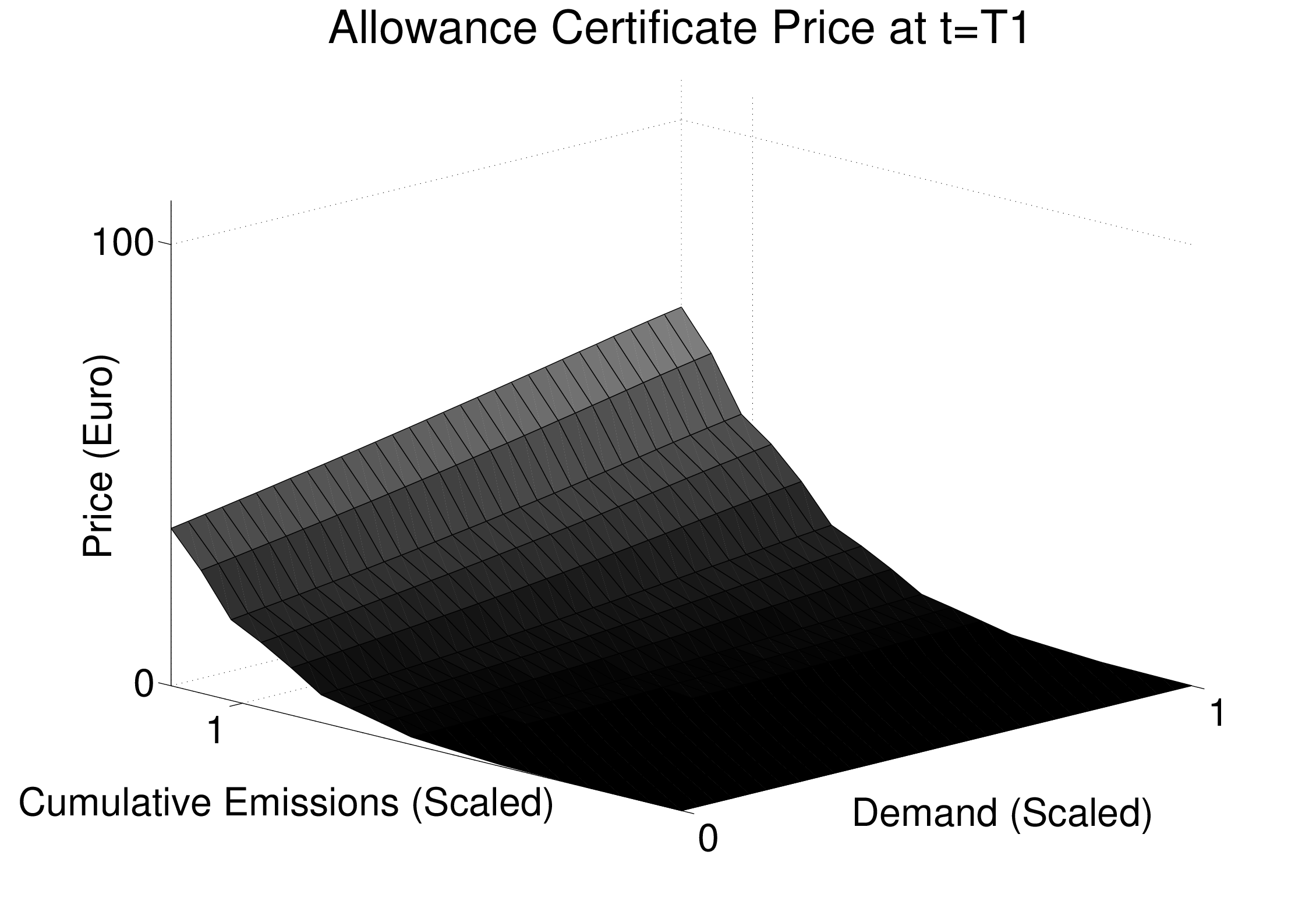}}
  \caption{The value of the first-period allowance certificate $(A^1_t)_{t\in[0,T_1]}$ at different times up to compliance, with two compliance
periods, connected by the borrowing, banking, and withdrawal mechanisms.}
\label{fig:allvalsurface_two_cp_BBW}
\end{figure}

As in the previous section, we analyze the prices of allowance certificates in this market in the two-period 
setting (see Figure \ref{fig:allvalsurface_two_cp_BBW}). During the second compliance
period the problem is equivalent to the market that only uses the banking and withdrawal mechanisms as described in section
\ref{str:allowance_pricing_BW}. This is the case, because in both markets the effect on the aggregate supply of certificates during the second period
is the same. Suppose the market is in compliance at time $t=T_1$. Then, a number $(E_{\text{cap}}^1 - E^1)$ of certificates are banked to the second
period and added to $E_{\text{cap}}^2$, independently of whether borrowing is allowed or not. Otherwise, if the market is not in compliance at time
$t=T_1$, a number $\min(E^1 - E_{\text{cap}}^1, E_{\text{cap}}^2)$ of certificates are withdrawn from $E_{\text{cap}}^2$ (if borrowing is not allowed),
or the same number are brought forward and hence also deducted from $E_{\text{cap}}^2$ (if borrowing is allowed). Therefore, for $i=2$ we solve the
PDE \eqref{eq:allowance_PDE_multiple_cp} with terminal condition \eqref{eq:allowance_terminal_cnd_second_cp_BW} and obtain
$\alpha_2=\alpha_2(\cdot,\cdot,\cdot;E^1)$. Subsequently, we solve \eqref{eq:allowance_PDE_multiple_cp} for $i=1$ together with the terminal condition
$\phi_1$ given by \eqref{eq:allowance_terminal_cnd_second_cp_BW}, where $A^2_{T_1}=\alpha_2(T_1,D,0;E)$. The boundary condition at $E=E_{\max}$ is
given by \eqref{eq:allowance_PDE_boundary_cond_first_cp_BW}.

\subsection{The impact of cap-and-trade}
The raison d'\^{e}tre of any cap-and-trade scheme is to reduce emissions. More precisely, its aim is to incentivize sufficient load shifting throughout
the trading period for the cumulative emissions not to reach the cap. In our modeling set-up we illustrate the effectiveness of cap-and-trade by
calculating the expected cumulative emissions at the end of the compliance period for different levels of the penalty $\Pi$. Recall that the penalty
represents an upper bound for the allowance price; the case $\Pi=0$ corresponds to business-as-usual and increasing $\Pi$ to a gradually more
aggressive cap-and-trade scheme.

We simulate the cumulative emissions process $(E_t)$ using the Monte Carlo scheme explained in Appendix \ref{ap:mc}, choosing $D_0=0.7\xi_{\max}$. We
repeat this simulation for values of the penalty ranging from zero to $200$ and calculate the mean of $E_T$, denoted by $\hat{E}_T$. We note that for
the present purpose of analyzing the cumulative emissions, the simulation of demand should take place under the physical measure $\mathbb{P}$, which
is related to $\mathbb{Q}$ by the market price of demand risk. This measure can be accurately estimated from market data (see, for example,
\cite{aEydeland2003} for different approaches). In the absence of a detailed data analysis, however, we follow the time-honored approach of letting
the market price of demand risk be constant and equal to zero. Therefore, for the purposes of our simulation we work with the stochastic differential
equation \eqref{eq:demand_coefficients}.

\begin{figure}[htbp]
  \centering
  \includegraphics[width=0.6\textwidth]{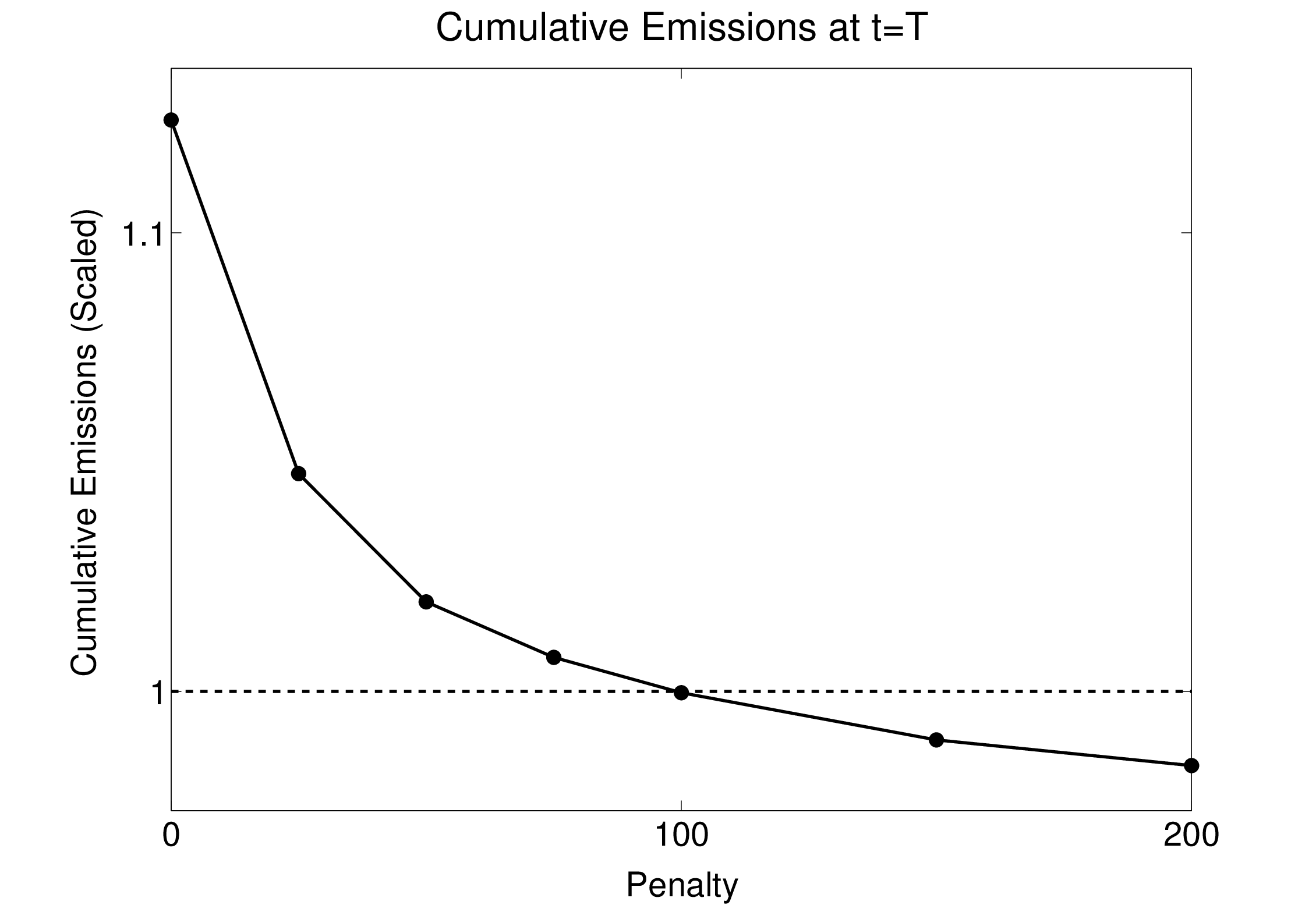}
  \caption{Mean of the cumulative emissions $E_T$ for different values of the penalty $\Pi$. The level of the cap is represented by the dashed line.}
 \label{fig:emsMC}
\end{figure}

Figure \ref{fig:emsMC} plots the results of the simulation of $10^6$ paths. Under business-as-usual the cumulative emissions are expected to exceed
the cap. As emissions trading is introduced, the market reacts by abating, but initially (at a penalty level of $\Pi=25$) the cumulative emissions still
exceed the cap. Further increases in the penalty (close to $\Pi=100$) lead to sufficient load shifting in order for the market to reach a state of
compliance. More aggressive regulation now leads only to small reductions in the cumulative emissions; i.e., our analysis confirms the well-known
stylized fact that emissions trading cannot incentivize firms to reduce cumulative emissions far below the cap.

\subsection{A call option on emissions}\label{str:call_num}
For the numerical solution of the call option, we specify
\begin{equation}\label{eq:call_PDE_bd2}
  v(t,D,E) = e^{-r(T-t)}\left(\Pi-e^{r(T-\tau)}K\right)^+ \quad \text{on } [0,\tau)\times(0,\xi_{\max})\times\{E=E_{\max}\}.
\end{equation}
The condition \eqref{eq:call_PDE_bd2} follows from recalling that when $E=E_{\max}$ the value of the allowance certificate $\alpha$ is given by
$\alpha(t,D,E)=e^{-r(T-t)}\Pi$. For the same reasons as put forward in section \ref{str:allowance_pricing_one_cp}, boundary conditions at
$D=0,\xi_{\max}$ and $E=0$ are not necessary.

\begin{figure}[htbp]
  \centering
  \subfloat[$t=\tau/2$.]{\label{fig:callvalsurf_one_cp_t1}\includegraphics[width=0.5\textwidth]{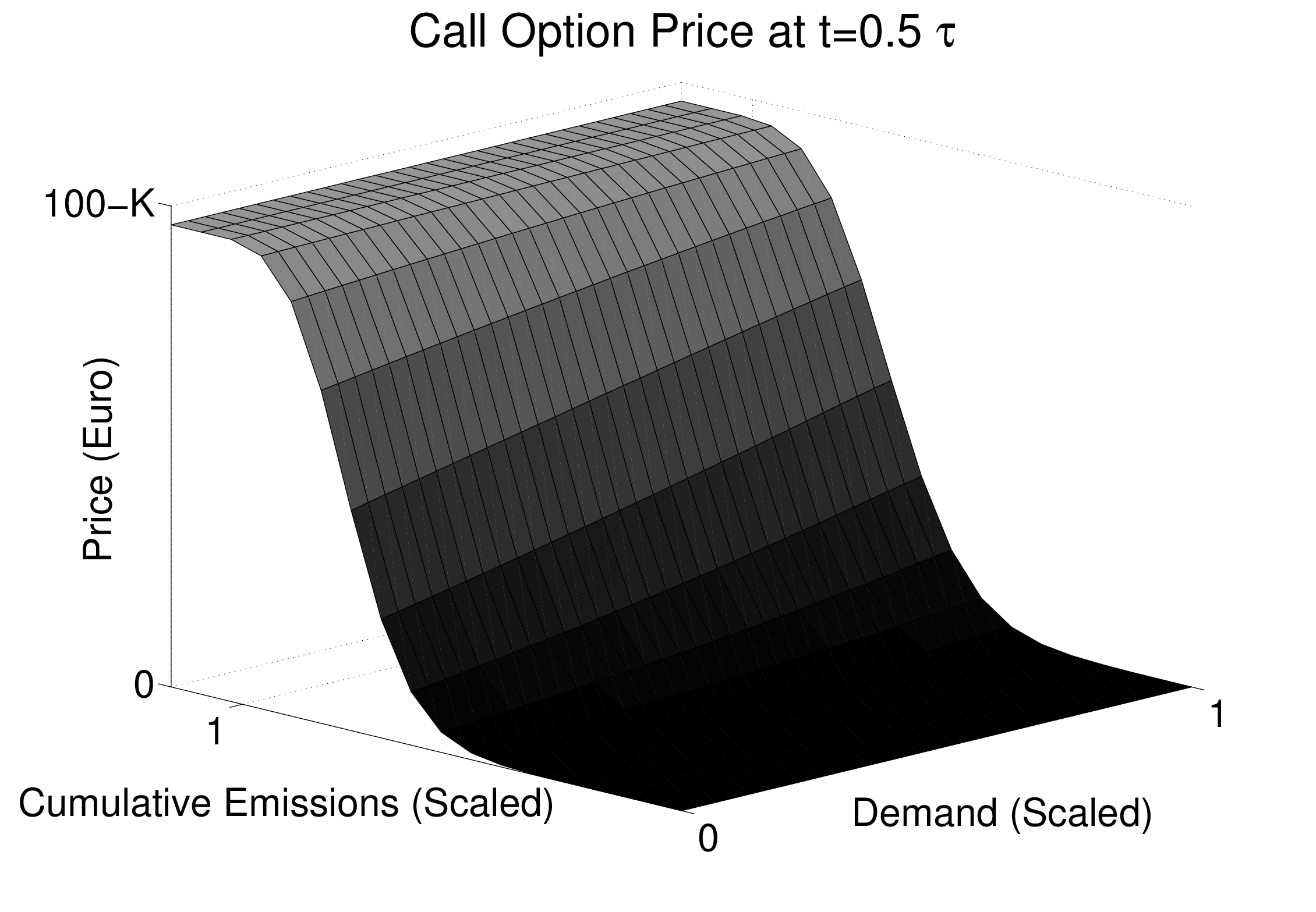}}
  \subfloat[$t=\tau$.]{\label{fig:callvalsurf_one_cp_tT}\includegraphics[width=0.5\textwidth]{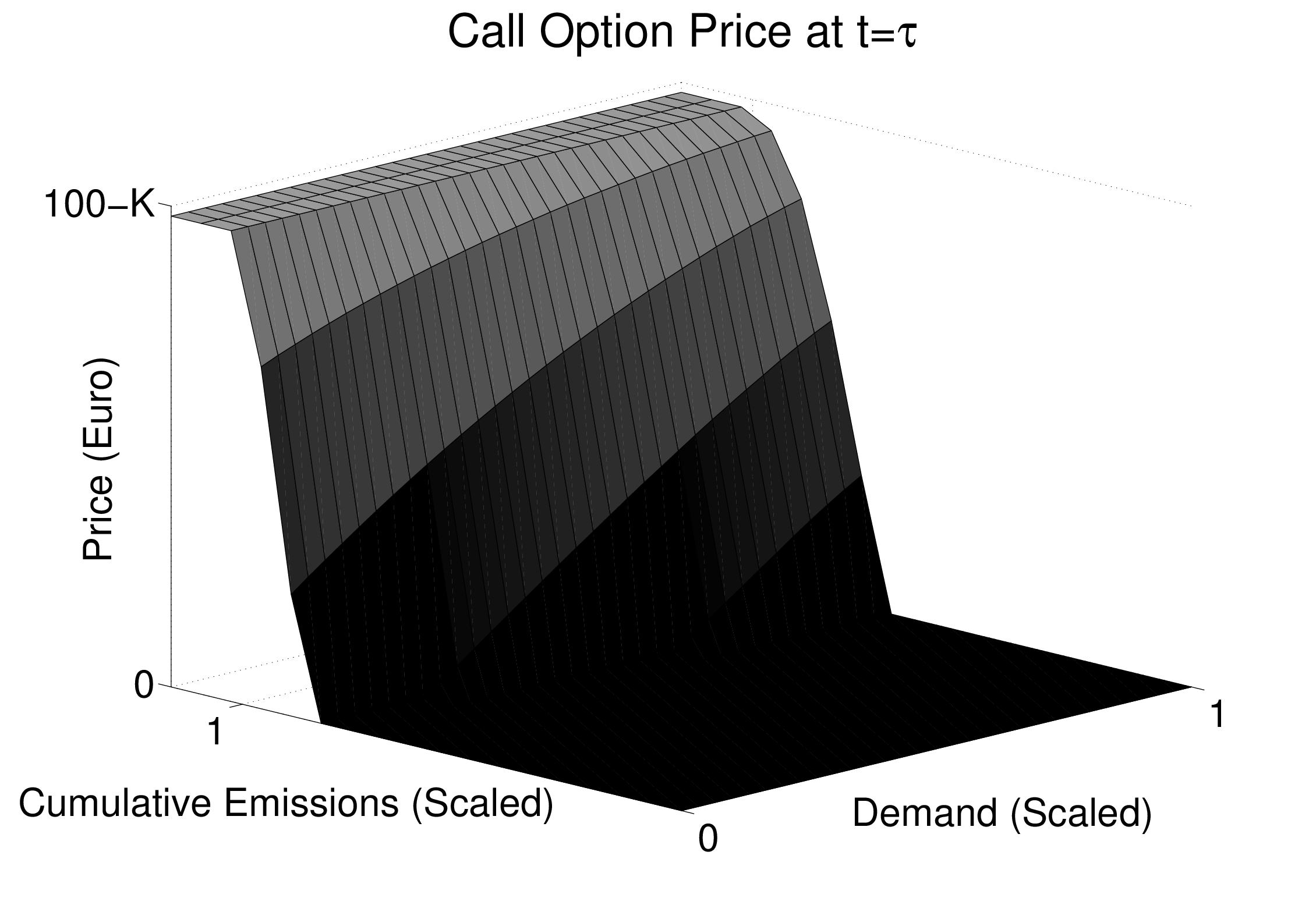}}
  \caption{The price of a call option, with strike $K$, on the allowance certificate, in an emissions market with one compliance period, at different
times up to expiry.}
 \label{fig:optionsurf_one_cp}
\end{figure}

Because the PDE \eqref{eq:call_PDE} requires the allowance price as an input parameter, it is necessary to solve
\eqref{eq:call_PDE} and \eqref{eq:allowance_PDE} in parallel in order to obtain the value of the option. The numerical scheme that determines the
price of the call as a function of the demand for electricity and the cumulative emissions is explained in Appendix \ref{ap:num_scheme_option_one_cp},
and the resulting value surface is plotted in Figure \ref{fig:optionsurf_one_cp}.

\section{Conclusion}
Emissions trading has become one of the most popular policy instruments employed by regulators to reduce global emissions. Allowance certificates---the 
key financial instruments in emission markets---and derivatives on them are traded actively on exchanges today despite the lack of an
established theoretical pricing framework taking the subtleties of these markets into account.

The contribution of this paper is threefold. First, by appealing to the bid stack, the key price-setting mechanism in electricity markets (which we
suggest is a reasonable description of more general markets as well), we introduce the idea of load shifting and show how the cost of carbon affects
which firms are supplying electricity to the market. This immediately allows us to deduce the rate at which emissions accumulate during a compliance
period. Because we derive the cumulative emissions process starting with an exogenously defined stochastic process for demand, we offer an explanation
for the noncompliance event, which is the main price driver of allowance certificates. Second, we embed the load shifting mechanism in a continuous
time pricing framework for allowance certificates, taking the form of a forward-backward stochastic differential equation (FBSDE). Within this set-up the
extension to price derivatives on the allowance certificate is immediate. Third, in the setting of an emission market with multiple compliance
periods, we analyze the impact different connecting mechanisms have on the allowance price.

\appendix

\section{Numerical solution of the allowance and option pricing problem}\label{ap:num_scheme_allowance}
We comment on the numerical scheme employed to solve the allowance and option pricing PDEs.

\subsection{One compliance period}\label{ap:num_scheme_one_cp}
We discretize the computational domain by choosing mesh widths $\Delta D$,
$\Delta E$ and a time step $\Delta t$. The discrete mesh points
$(D_i,E_j,t_k)$ are then defined by
\begin{align*}
D_i &:= i\Delta D,\\
E_j &:= j\Delta E,\\
t_k &:= k\Delta t.
\end{align*}
The finite difference scheme we employ produces approximations $\alpha^k_{i,j}$, which are assumed to converge to the true solution $\alpha$ as the
mesh width tends to zero.

Since the PDE \eqref{eq:allowance_PDE} is posed backward in time with a terminal condition, we choose a backward finite
difference for the time derivative in order to work with an explicit scheme.

In the $E$-direction we are approximating a conservation law PDE with discontinuous terminal condition. (For an in-depth discussion of numerical
schemes for these types of equations, see \cite{rLeVeque1990}.) The first derivative in the $E$-direction, relating to the nonlinear part of the PDE,
is discretized against the drift direction using a one-sided upwind difference. Because characteristic information is
propagating in the direction of decreasing $E$, this one-sided difference is also used to calculate the value of the approximation on the part of the
boundary corresponding to $E=0$.

In the $D$-direction the equation is parabolic everywhere except on the boundary, where it degenerates. Hence we use central differences to discretize
the first- and second-order derivatives. At the boundaries corresponding to $D=0$ and $D=\xi_{\max}$, where the second derivative vanishes and no
boundary conditions need to be specified, we again use a one-sided difference in our numerical scheme.
 
With smooth boundary data on a smooth domain and with a strictly decreasing (in $\alpha$) coefficient $\mu_E$, the scheme described above can be
expected to exhibit first-order convergence. In our setting, we expect the discontinuous terminal condition and the fact that $\mu_E$ is merely
decreasing in $\alpha$ to have adverse effects on the convergence rate.

We analyze the convergence of our numerical scheme in the supremum norm and the 1-norm at $t=0$; i.e., we calculate
\begin{equation*}
Err^\infty_l := \frac{ \|\alpha_{h_l} - \alpha_{h_{l+1}} \|_\infty}{\|\alpha_{h_l}\|_{\infty}} \quad \text{and} \quad Err^1_l := \frac{ \|\alpha_{h_l}
- \alpha_{h_{l+1}} \|_1}{\|\alpha_{h_l}\|_1},
\end{equation*}
where $\alpha_{h_l}$ represents the approximate solution given the vector of mesh parameters $h_l$ at refinement level $l$ and
\begin{equation*}
 \|\alpha^k_{i,j}\|_1:=\sum_{i,j} \left|\alpha^k_{i,j}\right| \Delta D \Delta E.
\end{equation*}
The parameters that define the different mesh widths are displayed in Table \ref{tab:num_mesh_param}; here we ensured that our choice
honors the Courant--Friedrichs--Lewy condition for the convergence of explicit schemes (cf.~\cite{rLeVeque1990}). As mentioned above, we expect the main
contributions to the error to stem from the hyperbolic part of the equation. Therefore, we choose a very fine grid in the $E$-direction for our
analysis.

\begin{table}[h]
\footnotesize
 \caption{Parameters for the convergence analysis of the numerical scheme.}\label{tab:num_mesh_param}
 \begin{center}
  \begin{tabular}{c|ccccc}
    \toprule
    & $h_1$ & $h_2$ & $h_3$ & $h_4$ & $h_5$\\
    \midrule
     $D_{\max} /\Delta D$ & 6 & 12 & 24 & 48 & 96\\
     $E_{\max} /\Delta E$ & 100 & 200 & 400 & 800 & 1600\\
     $1/\Delta t$ & 110 & 440 & 1760 & 7040 & 28160\\
    \bottomrule
  \end{tabular}
 \end{center}
\end{table}

Table \ref{tab:num_err} displays the results from our convergence study. Note that the error decays much faster in the 1-norm. This is not at all
surprising, as we expect the error from the discontinuous terminal condition to propagate in the direction of decreasing $E$. This leads to a
significant error on a small part of the grid, which is picked up by the infinity norm, whereas the approximation converges much faster everywhere
else, as shown by our analysis of the error in the 1-norm.

\begin{table}[h]
\footnotesize
 \caption{Numerical error in the supremum and the $1$-norm.}\label{tab:num_err}
 \begin{center}
  \begin{tabular}{c|cccc}
    \toprule
    $l$ & $1$ & $2$ & $3$ & $4$\\
    \midrule
     $Err^\infty_l$ & 0.0746 & 0.0355 & 0.0227 & 0.0105\\
     $Err^1_l$ & 0.0066 & 0.0020 & 0.0013 & 0.0006\\
    \bottomrule
  \end{tabular}
 \end{center}
\end{table}

\begin{figure}[htbp]
  \centering
  \includegraphics[width=0.7\textwidth]{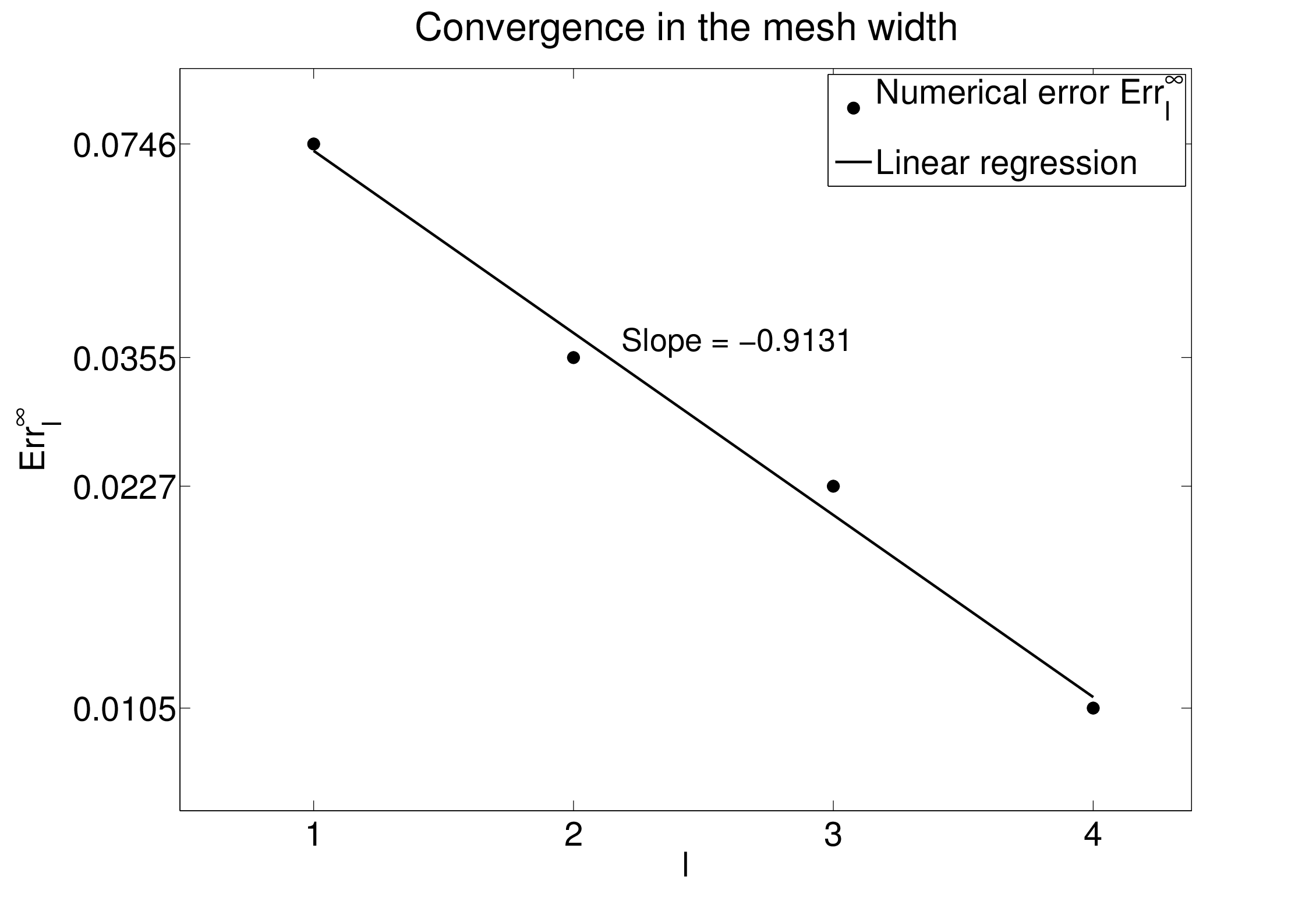}
  \caption{Illustration of the convergence of the finite difference scheme used to solve \eqref{eq:allowance_PDE}.}
 \label{fig:convergence}
\end{figure}

Figure \ref{fig:convergence} plots the error $Err^\infty_l$ as a function of the mesh width. From the slope of the line of best fit through the error
points, we estimate the convergence rate of our scheme to be $0.9131$ in the infinity norm.

\subsection{Multiple compliance periods}\label{ap:num_scheme_multiple_cp}
To deal with the path-dependency that enters the pricing problem in an emissions market with multiple compliance periods---in the case of banking
and withdrawal---through the terminal condition \eqref{eq:allowance_terminal_cnd_multiple_cp_BW} and \eqref{eq:allowance_terminal_cnd_second_cp_BW},
we have introduced the extra variable $E^1$.

The problem is then solved backward beginning with the second compliance period, for which we solve the PDE
\eqref{eq:allowance_PDE_multiple_cp} with corresponding boundary conditions and the terminal condition \eqref{eq:allowance_terminal_cnd_second_cp_BW},
introducing an extra dimension for the variable $E^1$. We then store $\alpha_2(T_1,D,0;E^1)$.

Subsequently, we solve the PDE corresponding to the first compliance period, using the stored values of $\alpha_2$ for the
evaluation of the terminal condition.

We compute the numerical approximation to each $\alpha_i$ using the scheme described in section \ref{ap:num_scheme_one_cp}. For a market in which banking,
borrowing, and withdrawal are implemented, the terminal condition at the end of the first period is modified in the obvious way.

We note that, compared with the single-period problem, the multiperiod problem has an extra dimension due to the variable $E^1$. When solving the
problem beginning with the second compliance period, this increases the complexity in two ways. First, the PDE \eqref{eq:allowance_PDE_multiple_cp} 
must be solved for a sufficiently large number of values of $E^1$. Second, we need to store $\alpha_2\left(T_2,D,0\right)$ since it is needed for 
the terminal condition of $\alpha_1$.

\subsection{Option pricing problem}\label{ap:num_scheme_option_one_cp}
We use the obvious modification of the earlier scheme described in section \ref{ap:num_scheme_one_cp} to solve the option pricing PDE \eqref{eq:call_PDE}.

\section{Monte Carlo simulation of cumulative emissions}\label{ap:mc}
Let $(D_k,E_k,A_k)$ denote the discrete time approximation to the FBSDE solution $(D_t,E_t,A_t)$ on the time grid $0 < \Delta t < 2\Delta t < \cdots <
n_k\Delta t = T$. At each time step we calculate $A_k$ by interpolating the discrete approximation $\alpha_{i,j}^k$ at $D_k,E_k$, beginning with the
initial values $D_0=d, E_0=0$. The approximations $(D_k,E_k)$ are obtained using a simple Euler scheme (cf. \cite{pGlasserman2004}). The discretized
version of $(D_t)$ is forced to be instantaneously reflecting at the boundaries $D_k=0$ and $D_k=\xi_{\max}$.

Using this discretization we simulate $n_c$ paths and, as usual, calculate the mean cumulative emissions $\hat{E}_T$, given by
\begin{equation*}
 \hat{E}_T: = \frac{1}{n_c}\sum_{i=1}^{n_c}E^i_{n_k},
\end{equation*}
where $E^i_{n_k}$ denotes the outcome of the simulation of the $i$th path. The corresponding standard error $\hat{\sigma}_{\hat{E}}$ is obtained by
\begin{equation*}
\hat{\sigma}_{\hat{E}}:= \sqrt{\frac{1}{n_c\left(n_c-1\right)}\sum_{i=1}^{n_c}\left(E^i_{n_k}-\hat{E}_T\right)^2}.
\end{equation*}
Table \ref{tab:num_MC} displays the results of our Monte Carlo simulation. 

\begin{table}[h]
\footnotesize
 \caption{Monte Carlo estimate of the mean cumulative emissions $\hat{E}_T$ and the corresponding standard error
$\hat{\sigma}_{\hat{E}}$.}\label{tab:num_MC}
 \begin{center}
  \begin{tabular}{c|ccccccc}
    \toprule
    $\Pi$ & $0$ & $25$ & $50$ & $75$ & $100$ & $150$ & $200$\\
    \midrule
     $\hat{E}_T$ ($1\times 10^8$) & 1.32 & 1.23 & 1.20 & 1.18 & 1.17 & 1.16 & 1.15\\
     $\hat{\sigma}_{\hat{E}}$ ($1\times 10^3$)  & 5.91 & 7.30 & 6.20 & 5.53 & 5.20 & 4.56 & 4.36 \\
    \bottomrule	
  \end{tabular}
 \end{center}
\end{table}


\subsection*{Acknowledgments}
The authors wish to thank Ren\'{e} Carmona, Michael Coulon, Jeff Dewynne, Dmitry Kramkov, and Christoph Reisinger for helpful comments.


\end{document}